\pdfoutput=1
  \RequirePackage{fix-cm}
  \documentclass[twocolumn]{svjour3}          
  \smartqed  
  \usepackage{graphicx}
  
  \usepackage{hyperref}
  \usepackage[style=numeric, backend=biber]{biblatex}
  
  \AtEveryBibitem{\clearfield{url}}

  \usepackage{amssymb}
  \usepackage{amsmath}
  \usepackage{array}
  \usepackage{float} 
  \usepackage{tikz}
  \usepackage{booktabs}
  \usepackage{multirow}
  \usepackage{arydshln} 
  \usepackage{lipsum}

  \usetikzlibrary{calc, positioning, shapes, arrows, decorations.pathmorphing}
  
\begin{document}
  
  \title{Revisiting Data Analysis with Pre-trained Foundation Models
  }
  \subtitle{ Pre-trained Foundation Model-Enhanced Scalability}
  
  
  \author{Chen Liang         \and
          Donghua Yang  \and
          Zheng Liang   \and 
          Zhiyu Liang   \and \\
          Tianle Zhang \and
          Boyu Xiao \and  
          Yuqing Yang \and  
          Wenqi Wang \and  \\
          Hongzhi Wang$\ast$ \thanks{$\ast$ Hongzhi Wang is the corresponding author.}
  }
  
  
  \institute{ 
  Chen Liang, \email{23B903050@stu.hit.edu.cn} \and \\
  Donghua Yang, \email{yang.dh@hit.edu.cn} \and \\
  Zheng Liang, \email{lz20@hit.edu.cn} \and \\ 
  Zhiyu Liang, \email{zyliang@stu.hit.edu} \and \\
  Tianle Zhang,  \email{2021112956@stu.hit.edu.cn} \and\\
  Boyu Xiao, \email{2022110524@stu.hit.edu.cn} \and \\
  Yuqing Yang, \email{2022111927@stu.hit.edu.cn} \and \\
  Wenqi Wang, \email{107489194@qq.com} \and \\
  Hongzhi Wang, \email{wangzh@hit.edu.cn} \at 
                Harin Institute of Technology, Harbin, China  \\
  }
  
  \date{Received: date / Accepted: date}

  \maketitle
  
  \begin{abstract}
  Data analysis focuses on harnessing advanced statistics, programming, and machine learning techniques to extract valuable insights from vast datasets. An increasing volume and variety of research emerged, addressing datasets of diverse modalities, formats, scales, and resolutions across various industries. However, experienced data analysts often find themselves overwhelmed by intricate details in ad-hoc solutions or attempts to extract the semantics of grounded data properly. This makes it difficult to maintain and scale to more complex systems. Pre-trained foundation models (PFMs), grounded with a large amount of grounded data that previous data analysis methods can not fully understand, leverage complete statistics that combine reasoning of an admissible subset of results and statistical approximations by surprising engineering effects, to automate and enhance the analysis process. It pushes us to revisit data analysis to make better sense of data with PFMs. This paper provides a comprehensive review of systematic approaches to optimizing data analysis through the power of PFMs, while critically identifying the limitations of PFMs, to establish a roadmap for their future application in data analysis.

  \end{abstract}
  
  \section{Introduction}
  Data analysis aims to identify and understand objects, their evolution trend, and relationships between them~\cite{3Wmodel,nisbet2009handbook}, promoting the ability to solve real-world problems. The scope of data analysis branches from pattern mining, and retrieval to prediction and causal inference~\cite{pearl2018book}. It serves as the cornerstone for finding linkage between diverse phenomena and underlying principles, mechanisms, or causes, driving informed decision-making across diverse domains~\cite{provost2013data}. As techniques emerge in various applications, it enable organizations to anticipate optimizations for operations, tailor services to meet evolving human needs, and make scientific discoveries. 
  
  Though previous methods can propose ad-hoc solutions by imperatively integrating human knowledge, they can not make full sense of available data and general computation. They are not designed but expected to thoroughly extract grounded semantics implied in data. As a successful practice of knowledge engineering, PFM facilitates the identification of grounded concepts and the discovery of new phenomena. This urges us to revisit data analysis with a universal and thorough perspective to make full use of PFMs for new findings.
  
  \subsection{Tasks and Challenges in Scaling of Data Analysis}\label{sec:skeleton}
  
  \begin{figure*}[h] 
    \centering 
    \includegraphics[width=\textwidth]{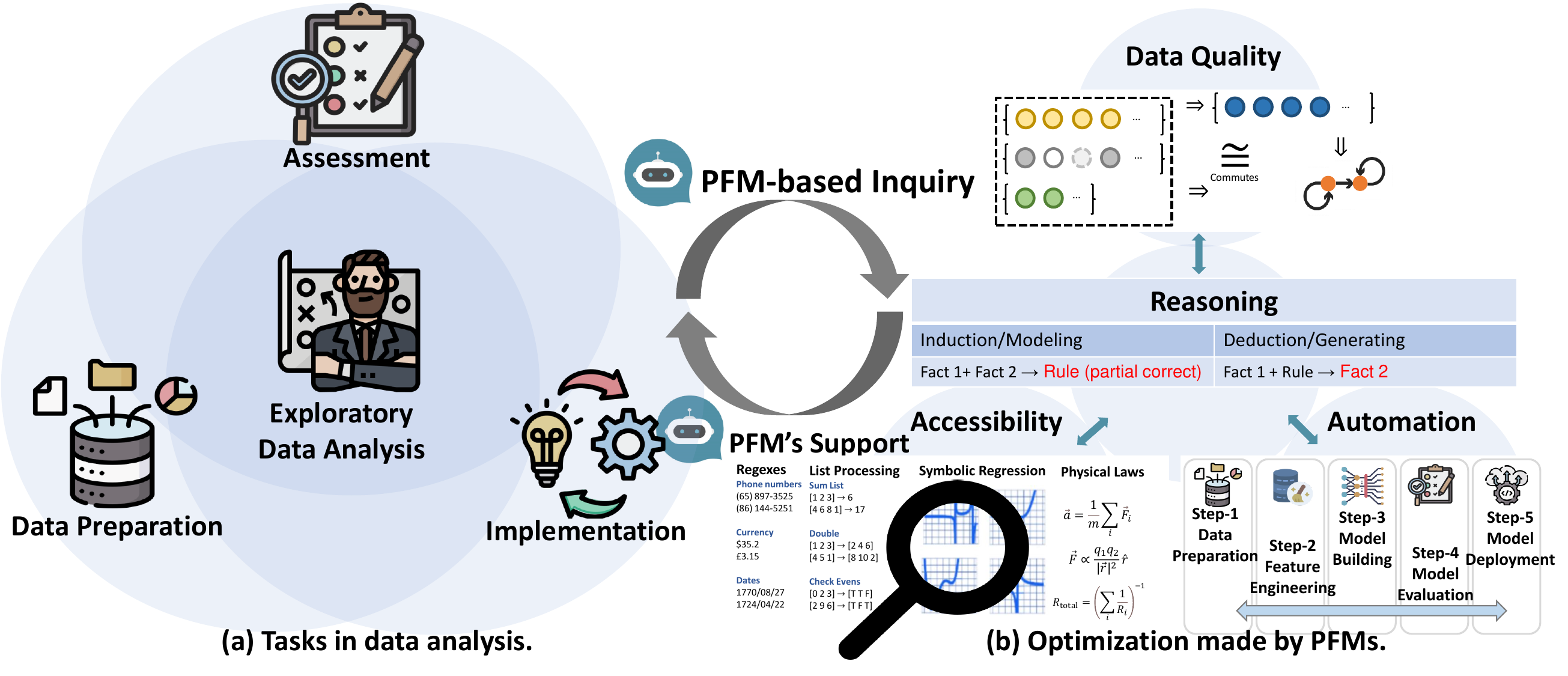} \label{fig:overview}
  
  \caption{\textbf{A framework for data analysis tasks and challenges.} (a) Data preparation: handling data preparation and ensuring quality for analysis. Exploration: facilitating interactive analysis to uncover trends and patterns. Implementation: applying specific methods for reasoning, modeling, and decision-making. Assessment: validating results and ensuring reliability. These tasks interact with and overlap with each other. (b) PFMs address the challenges of accessibility, quality optimization, and automation, powered and regularized by PMF-enhanced reasoning, which in turn enable the effective execution of these data analysis tasks. These applications and challenges are detailed in \S~\ref{sec:methods}. PFMs commute specific tasks with the essence of intelligent ability provided by previous tools and theories.}
  \end{figure*}
  
  We first introduce the scope and logical skeleton of our revisitation of data analysis — tasks that cover most of the needs of data analysis~\cite{launer2014modern, brandt1976statistical, van2020data,olson2003data, myatt2007making, brooker2020practical} and challenges an analyst will meet that are easier when benefiting from PFMs. In this section, we overview the specific topics that form the overall structure of this paper.

  Data analysis techniques originate from multiple application fields~\cite{SPIREX, vertsel2024hybrid,gerussi2022llm,Zhang2024LargeLM,truhn2023large}. Though loosely grouped, they have the same ability to find useful structures in all kinds of data records~\cite{launer2014modern}. As branches of this experimental science~\cite{brandt1976statistical}, several well-established utilitarian tasks fall within the scope of data analysis: (\S~\ref{sec:data_management}) data preparation~\cite{van2020data,olson2003data}, which transform unorganized datasets with quality issues to high usable ones for data analysis, (\S~\ref{sec:explore}) exploratory data analysis~\cite{myatt2007making}, which helps analysts to interact with highly complex databases and tools which they are not familiar with, (\S~\ref{sec:da_methods}) implementation of data analysis methods~\cite{brandt1976statistical,nisbet2009handbook}, which imperatively implement data analysis tasks with primary operations, and (\S~\ref{sec:correctness}) assessing analyzed results~\cite{brooker2020practical,bruce2020practical,kenett2016information}, which assesses the reliability and practicability of analytical conclusions. We organize their definitions, overview them within the scope of our topic, and related surveys about applications of PFMs in \S~\ref{sec:task_solving}.
  
  Although PFMs can be introduced in alignment with these tasks, it is important to further identify the specific challenges that hinder the seamless execution of complex analytical tasks, which PFMs can help overcome more effectively. In this paper, we revisit these tasks with emphasis on introducing PFMs and then infer a comprehensive and concise summary of the key challenges, which are intuitively overviewed in Fig.~\ref{fig:overview}, and their subordinate sub-challenges. These challenges are related to the systematic optimization of data analysis through PFMs, each of which consists of several sub-challenges. The sections (\S) correspond to where PFMs can address these challenges and drive optimization.

  \begin{table*}[ht]
    \centering
    \caption{Overview of PFMs enhanced systematic optimization methodologies.}
    \label{tab:structure_citations}
    \resizebox{1\textwidth}{!}{
    \begin{tabular}{llllll}
    \toprule
  \multirow{24}{*}{PFMs Based Optimizations (\S~\ref{sec:methods})} 
  & \multirow{8}{*}{Reasoning (\S~\ref{sec:dsl})} 
    & \multirow{3}{*}{Representation learning (\S~\ref{sec:representation})}
    &  &  & \cite{buchholzrobustness, feyposition, yuan2023power, blaauwbroekgraph2tac, HansenCF24, DeletangRDCGMGW24, GrandWBOLTA24} \\ & & & & & \cite{zhengprise, KoziolekGHALE24, ahuja2023interventional, jain2024r2e,LLMschema,TURL} \\
    & & & & & \cite{FMcard} \\
  \cdashline{3-6}
  &  & \multirow{3}{*}{Domain Specific Logic (\S~\ref{sec:reasoning})} 
    & \multirow{2}{*}{Induction} 
    &  & \cite{ye2024satlm, jha2023counterexample, ni2024iterclean, qi2024cleanagent, ma2023insightpilot, Dibia2023LIDAAT} \\ & & & & & \cite{Ellis2020DreamCoderGG, Tang2024WorldCoderAM,li2023resdsql, gu2023few, vos2022towards, deem} \\
  \cdashline{4-6}
  &  &  & Deduction 
    &  & \cite{Li2024IsPB, Khakhar2023PACPS, qi2024cleanagent, Pei2023CanLL, Wang2024TheoremLlamaTG, wang2023large} \\
  \cdashline{3-6}
  &  & \multirow{3}{*}{Consolidation (\S~\ref{sec:consolidate_dsl})} 
    & Reusability
    &  & \cite{jain2024r2e, repocomp, GrandWBOLTA24, constructKG, KGobjectrecognition,vectorstorage} \\
  \cdashline{4-6}
  &  &  & Understandable 
    &  & \cite{ma2023insightpilot, Dibia2023LIDAAT} \\
  \cdashline{4-6}
  &  &  & Thoroughness 
    &  & \cite{brand2023parameterized, Wang2024TheoremLlamaTG, Carrott2024CoqPytPN} \\
  \cdashline{2-6}
  & \multirow{9}{*}{Accessibility (\S~\ref{sec:interpretability})} 
    & \multirow{6}{*}{Interface (\S~\ref{sec:interface})} 
    & \multirow{3}{*}{Data of Concern} 
    &  & \cite{li2023resdsql, gu2023few, cheng2022binding, huang2023kosa, li2024flexkbqa, Lehmann2023LanguageMA,LLMschema} \\ & & & & & \cite{readyforNL2SQL,smalllargemodelNL2SQL,text2sqlevaluation,GPTuner,ReActTable,NLPTune} \\
    & & & & & \cite{CatSQL, trummer2022codexdb,NeuralDB} \\
  \cdashline{4-6}
  &  &  & Methods of Concern 
    &  & \cite{lai2023ds, zhu2024retrieval, ngom2024mallet} \\
  \cdashline{4-6}
  &  &  & Documenting/Translation 
    &  & \cite{cummins2023large, cummins2024meta, li2024can, GrandWBOLTA24, Bavishi2022NeurosymbolicRF} \\
  \cdashline{4-6}
  &  &  & Interaction 
    &  & \cite{lai2023ds, li2024can} \\
  \cdashline{3-6}
  &  & \multirow{3}{*}{Interpretability/Editability (\S~\ref{sec:editability})} 
    & Balancing Expressiveness 
    &  & \cite{reizingerposition, nam2024using, singh2023augmenting, ko2024filling, zheng2024revolutionizing} \\
  \cdashline{4-6}
  &  &  & \multirow{2}{*}{Balancing Stability}
    &  & \cite{vojivr2020editable, popov2019neural, grinsztajn2022tree, reizingerposition, nam2024optimized, caruana2022data} \\ & & & & & \cite{vertsel2024hybrid, gerussi2022llm, Zhang2024LargeLM} \\
  \cdashline{2-6}
  & \multirow{4}{*}{Data Quality (\S~\ref{sec:data_quality})} 
    & \multirow{2}{*}{Sampling/Generating (\S~\ref{sec:sample_generating})} 
    & Active Sampling 
    &  & \cite{Wang2023SoloDD,LLMschema,tabulargeneration,CHORUS} \\
  \cdashline{4-6}
  &  &  & Deductive Generating 
    &  & \cite{Nobari2023DTTAE, Loem2023SAIEFS, huamortizing, du2024enhancing, weng2023g} \\
  \cdashline{3-6}
  &  & \multirow{2}{*}{Robustness (\S~\ref{sec:robustness})} 
    & Data Cleaning. 
    &  & \cite{li2024towards, peng2022self} \\
  \cdashline{4-6}
  &  &  & Robust Methods 
    &  & \cite{Miao2022LearningIA, Neu2022GeneralizationBV, Atzeni2023InfusingLS, 094dai2017good, 097yoon2017semi} \\
  \cdashline{2-6}
  & \multirow{2}{*}{Automation (\S~\ref{sec:auto_ml})} 
    & Consolidating AutoML (\S~\ref{sec:consolidate_automl}) 
    &  &  & \cite{Hollmann2023LargeLM, sayed2024gizaml, liu2023jarvix, bai2024transformers} \\
  \cdashline{3-6}
  &  & Scaling AutoML (\S~\ref{sec:scaling_automl}) 
    &  &  & \cite{SongY00024, HsuMTW23, 009brown2020language, EoTGE, reizingerposition, cheng2022binding} \\
  \bottomrule
    \end{tabular}
    }
\end{table*}

  \paragraph{Scalable Reasoning (\S~\ref{sec:dsl})}  
  As data analysis advances, there are growing concerns about the limits of what it can handle. With the increasing volume and variety of data, obsolete analysis methods struggle to keep up, especially when dealing with data in different formats, sizes, and resolutions. Some of the main challenges include whether existing knowledge can be applied to more complex systems, and how to keep analytical methods up to date.
  
  \underline{(1) Representation learning of concepts.} This involves capturing links between real-world concepts and terms in an analysis problem, helping to clarify the meaning and semantics of terms within the problem's context. Representation learning maps latent structures in data to geometric signatures or symbolic representations that identify concepts~\cite{ahuja2023interventional}. Structural methods scale better for more complex systems, due to their high maintainability.
  
  \underline{(2) Reasoning with domain-specific concepts.} At the same time, data analysis methods must be continuously expanded to keep up with the fast-changing demands of analysis. Reasoning powers algorithms that can extract concepts from data samples~\cite{brand2023parameterized} and generate new insights.
  
  \underline{(3) Consolidation of domain-specific language.} Fragmented approaches can hinder consistency and reproducibility, making it hard to establish common benchmarks. Consolidating methods is essential to encourage collaboration, and ensure the maintainability of high-quality solutions with high generalization power and conciseness, which can improve the reusability and reliability of results across different industries.
  
  \paragraph{Accessibility of Datasets and Models (\S~\ref{sec:interpretability})}  
  While the effectiveness and scope of data analysis continue to grow, the operability of complex models remains a significant challenge. Data analysts prefer more accessible methods to avoid specific risks and inject or delete prior knowledge in these models. 
  
  \underline{(1) Data analysis interface.} Improving interfaces is essential to ensure that human intervention and alignment are both efficient and effective. It further aligns the intention and grounded goal of data analysts with the provided data, methods, and models.

  \underline{(2) Interpretability and editability.} Although advanced models can provide valuable insights, their "black-box" nature often hides the decision-making process. This lack of transparency makes it difficult to validate hypotheses, communicate results to non-technical stakeholders, and build trust in the findings.

  \paragraph{Data Quality Issues (\S~\ref{sec:data_quality})}  
  Data quality directly impacts the reliability of the conclusions drawn from data analysis. Data quality issues involve consistency, accuracy, completeness, timeliness, and identity of the same concept. They can be more concisely concluded with consistency and completeness, assuming that finally acquired data is up to date and concepts can be identified among multiple different data sources according to consistent rules with different data sources.
  
  \underline{(1) Representative data sampling/generating.} Completeness involves addressing missing values, gaps, and incomplete records that can arise during data collection and integration. Sampling representative samples and generating missing data deduced from inferred rules and principles help to improve the completeness of inducing knowledge from data. 
  
  \underline{(2) Robustness against errors.} Robustness is related to the consistency of datasets and models, which refers to the absence of contradictions, ensuring that integrated heterogeneous data sources adhere to compatible formats and standards. Achieving consistency requires extensive preprocessing and harmonization to create unified datasets or acquire a model that is robust to inconsistent datasets. Without robust management of consistency and completeness, analytical outcomes are compromised, potentially leading to misinformed decisions and reduced stakeholder trust.
  
  \paragraph{Automation of Data Analysis (\S~\ref{sec:auto_ml}
  )} Traditionally, the above challenges have been tackled through manual efforts and ad-hoc solutions for various data analysis tasks. For application sectors where human interference is not always available, self-disciplined decision plans should be made with acquired data. In the context of machine learning, automated machine learning (AutoML) refers to the automated search for highly effective machine learning methods. Philosophically, machine learning involves inducing concepts from finite samples within a hypothesis space. Thus, we consider AutoML to be the most appropriate term for the automated search for data analysis methods and models. 
  
  \underline{(1) Consolidation of AutoML.} The first challenge is to enhance the performance of AutoML within the existing constraints of the hypothesis class. Specifically, the primary concern within the AutoML community is improving both the efficiency and effectiveness of AutoML. 
  
  \underline{(2) Scaling AutoML.} Additionally, extending and relaxing the effective scope of AutoML with structural guidance presents a challenge. It is more meaningful to introduce a more scalable and human-centered AutoML. For example, domain experts may input their knowledge into the scaled system to configure more suitable data analysis methods. This would make AutoML more meaningful for solving more technical and mathematical problems when data scientists have deeper needs.

  \subsection{Introducing PFMs into Data Analysis} 
  
  While (Pre-trained Language Models) PLMs can only operate on a sequence of tokens or encoded latent embeddings, other foundation models can operate on various multi-modal data which also preserve algebraic or statistic structures, formalizable logic clauses, and algorithms. Foundation models and pre-trained foundation models (PFMs) may refer to all the above models which benefit from pertaining large amounts of parameters with enormous datasets to perform all kinds of reasoning and decision tasks. Taxonomy for the PFMs according to their data modality, model architecture, and training methods are well organized in specialized surveys~\cite{YangJTHFJZYH24,zhou2023comprehensive,LiangWNJ0SPW24,videounderstanding,yin2023survey,wu2023multimodal}. 
  
  In this paper, we dig deeper into why, and how PFMs can systematically strengthen the utility and scalability of data analysis. We conclude that PFMs have several advantages that can benefit data analysis. 
  
  \paragraph{Ability to name and understand symbolic and formal concepts.} PFMs provide opportunities for operating formal logical systems. Separated from other machine learning methods, the combination of operating with logical systems of different modalities, completeness, and consistency is the key advantage of PFMs. Their pretext tasks (e.g. next-world prediction on multiple corpora) can represent manipulations of a variety of data~\cite{yuan2023power}. PFMs pre-trained by large amounts of text and code can understand and compose complex semantics and concepts defined by nested or recursively built syntax structures~\cite{phi1}. When processing a wide range of objects, PFMs can approximately correctly choose operations and morphisms to build complex concepts~\cite{009brown2020language}. 
  
  For example, a neural symbolic system based on PFMs can help analysts construct a formalized representation of a problem, helping them to mitigate or ignore complexities of syntax and semantic checking~\cite{Bavishi2022NeurosymbolicRF}. Code generation systems, e.g., Copilot, provide different modes with high usability to help programmers implement analysis programs~\cite{Barke2022GroundedCH}. They can not only understand declarative high-level programming language and imperative low-level languages but are also designed to compose them correctly according to human instructions and intentions. 
  
  \paragraph{Composing probably approximately correct (PAC) hypothesis during in-context learning.} PFMs can perform multiple tasks simply with brief demonstrations and a few samples without training and fine-tuning. Each of those inference steps or chains can be seen as an entire approximate correct identification of concepts. Primary knowledge or common sense, which is frequently presented in pre-training datasets, is induced by the structure of datasets and approximated by the data distribution. This primary knowledge is useful for compressing general-purpose concepts and hypotheses. 
  
  In \S~\ref{sec:interpretability}, we introduce concepts class, learning algorithms, hypothesis class, and data distribution in the languages of probably approximately correct (PAC) learning theory, which can recover the most purposes of induction in data analysis. This will help us to divide intrinsic methods and research directions that introduce the power of PFMs into several categories. We will see how the composition of multidomain operators and concepts helps to introduce multidimensional optimization into the data analysis sector.
  
  The purpose of composing concepts and hypotheses in a data analysis problem is to separate specific runtime implementation of methods with the goal of analysis, which is the core of scaling and optimization of data analysis. In this process, PFMs are proven to be playing a significant part. 
  
  \paragraph{Generalization across different modalities and categories.} Foundation models can generate and make judgments about unseen objects from the target category (e.g., images or videos~\cite{queryvideo}) using the structural information from the source category (e.g., texts)~\cite{Yuan23a}. This generalization ability can't be fully estimated by only discussing statistical generalization~\cite{ReizingerUMKBH24}. Operations and laws in algebras, which are demonstrations of structures of logic systems, perform exportation in reasoning. Correct operation on these structures made by PFMs should be identified with disentangled data distribution and algebraic knowledge. The key factor of this type of generalization is the ability to compose maps between these structures of different modalities and categories~\cite{DuK24}.

  PFMs bring opportunities for systematic optimization of data analysis. Instead of introducing challenges with the order of life circle of PFMs, we are more concerned with completeness and thoroughness where PFMs help to achieve the purpose of data analysis. PFMs have shown multi-dimensional advantages in introducing systematic optimization into data analysis. Providing the capability of enhancing the automation of formal reasoning with multi-modal knowledge and integrating interdisciplinary knowledge, PFMs introduce various opportunities in the scope of data analysis. This paper is a systematic review of PFMs-empowered data analysis methodologies, our contributions are listed below.
  
  \begin{itemize}
      \item We provide a comprehensive organization of utilitarian data analysis tasks and thoroughly analyze the value proposition of incorporating PFMs in (\S~\ref{sec:task_solving}). We further focus on technical and in-depth aspects, such as reasoning, interpretability, and other key factors that enhance data analysis in (\S~\ref{sec:methods}).
      
      \item We not only broadly introduce applications of PFMs in data analysis in terms of key tasks and methodologies, but also dig into much deeper details of obstacles and theoretical analysis, making a sound evaluation of challenges for every application domain.
      
      \item We aim to extend data analysis with previous analytical results from PFMs, generalizing their application conditions and building a solid foundation for further innovations. We summarise state-of-the-art methodologies and findings, clarifying their (dis)advantages and providing a roadmap for challenges and further research.
  
  \end{itemize}

  \section{Data Analysis Tasks Solving Empowered by PFMs}\label{sec:task_solving}
  
  We first review challenges met by data analysis, they are presented in tasks including data management, exploratory data analysis (EDA), implementation data analysis, and assessment of analyzed results. We put stress on the complementarities of conventional methods and new techniques. We achieve this by pointing out the key purposes of data analysis, preserving the advantages of traditional theoretical results and smoothly introducing PFMs to boost performance and efficiency. This section formulate the applications and benefits of introducing PFMs.

  \vspace{1cm} 
  \newlength{\leftParentNodeWidth}
  \setlength{\leftParentNodeWidth}{3cm} 
  
  \newlength{\rightParentNodeWidth}
  \setlength{\rightParentNodeWidth}{2cm} 
  
  \newlength{\leftNodeWidth}
  \setlength{\leftNodeWidth}{2.8cm} 
  
  \newlength{\middleNodeWidth}
  \setlength{\middleNodeWidth}{4.2cm} 
  
  \newlength{\middleNodeHeight}
  \setlength{\middleNodeHeight}{0.2cm} 
  
  \newlength{\rightNodeWidth}
  \setlength{\rightNodeWidth}{3.4cm} 
  
  \newlength{\leftVerticalSpacing}
  \setlength{\leftVerticalSpacing}{0.9cm} 
  
  \newlength{\middleVerticalSpacing}
  \setlength{\middleVerticalSpacing}{0.6cm} 
  
  \newlength{\rightVerticalSpacing}
  \setlength{\rightVerticalSpacing}{0.9cm} 
  
  \newlength{\horizontalSpacing}
  \setlength{\horizontalSpacing}{13.5cm} 
  
  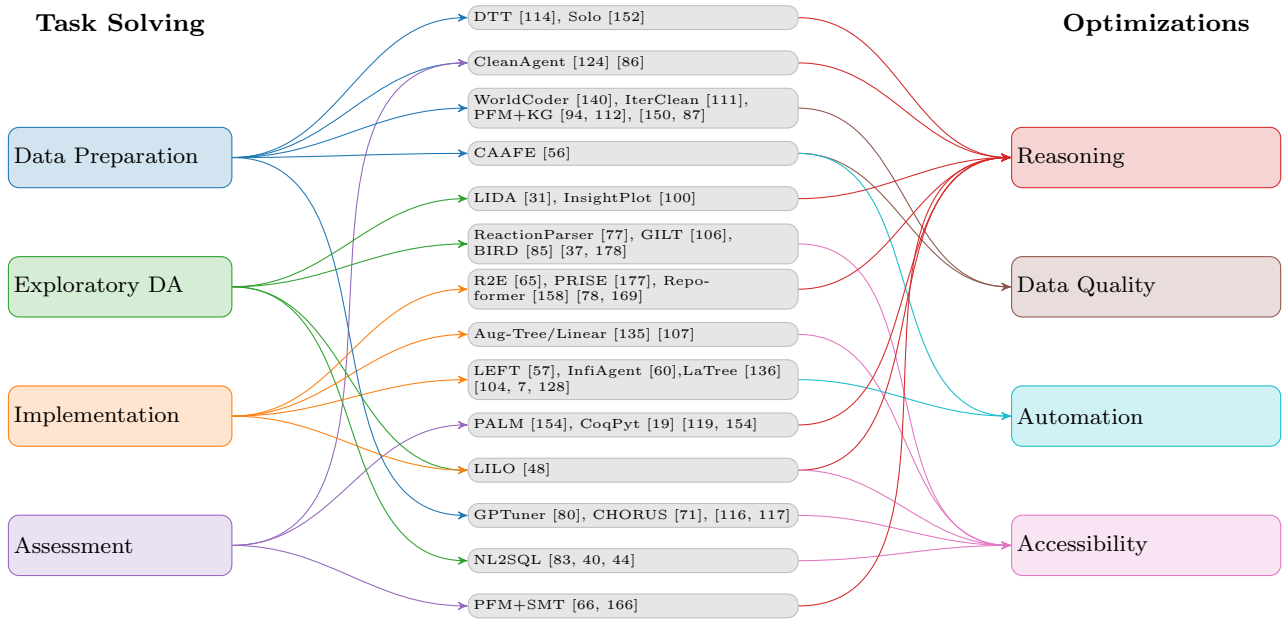
\begin{figure*}[h]

    \centering
\begin{tikzpicture}[
  auto,
  emptyNode/.style={
    draw=none,    
    fill=none,    
    minimum height=0.8cm,  
    inner sep=0pt,   
    align=center,
    scale=1.1
  },
  baseNode/.style={
    rectangle,
    draw,
    rounded corners,
    minimum height=0.8cm,
    inner sep=2pt,
    font=\small
  },
  leftParentNode/.style={
    emptyNode,
    text width=\leftParentNodeWidth,
    text=black,
    font=\bfseries,
  },
  rightParentNode/.style={
    emptyNode,
    text width=\rightParentNodeWidth,
    text=black,
    font=\bfseries
  },
  middleNode/.style={
    baseNode,
    text width=\middleNodeWidth,
    fill=gray!20,
    draw=gray!50,
    text=black,
    font=\tiny,
    minimum height=\middleNodeHeight
  },
  >=stealth'
]
  \definecolor{DataManagementColor}{RGB}{31,119,180} 
  \definecolor{ExploratoryDataAnalysisColor}{RGB}{44,160,44} 
  \definecolor{ImplementationColor}{RGB}{255,127,14} 
  \definecolor{AssessmentColor}{RGB}{148,103,189} 
  
  \definecolor{DSLUnderstandingColor}{RGB}{214,39,40} 
  \definecolor{DataQualityOptimizationColor}{RGB}{140,86,75} 
  \definecolor{AutoMLColor}{RGB}{23,190,207} 
  \definecolor{AccessibilityColor}{RGB}{227,119,194} 
  
  \node[leftParentNode] (left_parent) at (0,-2) {Task Solving};
  
  \node[rightParentNode] (right_parent) at ($(left_parent)+(\horizontalSpacing,0)$) {Optimizations};
  
  \coordinate (middleStart) at ($(left_parent)!0.5!(right_parent)$);
  
  \node[baseNode, fill=DataManagementColor!20, draw=DataManagementColor, text width=\leftNodeWidth, below=\leftVerticalSpacing of left_parent] (data_management) {Data Preparation};
  \node[baseNode, fill=ExploratoryDataAnalysisColor!20, draw=ExploratoryDataAnalysisColor, text width=\leftNodeWidth, below=\leftVerticalSpacing of data_management] (exploratory_data_analysis) {Exploratory DA};
  \node[baseNode, fill=ImplementationColor!20, draw=ImplementationColor, text width=\leftNodeWidth, below=\leftVerticalSpacing of exploratory_data_analysis] (implementing_methods) {Implementation};
  \node[baseNode, fill=AssessmentColor!20, draw=AssessmentColor, text width=\leftNodeWidth, below=\leftVerticalSpacing of implementing_methods] (assessing_results) {Assessment};
  

  \node[baseNode, fill=DSLUnderstandingColor!20, draw=DSLUnderstandingColor, text width=\rightNodeWidth, below=\rightVerticalSpacing of right_parent] (dsl_understanding) {Reasoning};
  \node[baseNode, fill=DataQualityOptimizationColor!20, draw=DataQualityOptimizationColor, text width=\rightNodeWidth, below=\rightVerticalSpacing of dsl_understanding] (PFM_quality) {Data Quality};
  \node[baseNode, fill=AutoMLColor!20, draw=AutoMLColor, text width=\rightNodeWidth, below=\rightVerticalSpacing of PFM_quality] (automated_ml) {Automation};
  \node[baseNode, fill=AccessibilityColor!20, draw=AccessibilityColor, text width=\rightNodeWidth, below=\rightVerticalSpacing of automated_ml] (accessible_models) {Accessibility};
  
  \def\middleNodes{
  {group1}/{DTT~\cite{Nobari2023DTTAE}, Solo~\cite{Wang2023SoloDD}},
    {group2}/{CleanAgent~\cite{qi2024cleanagent} \cite{Li2024IsPB}},
    {group3}/{WorldCoder~\cite{Tang2024WorldCoderAM}, IterClean~\cite{ni2024iterclean}, PFM+KG~\cite{KGobjectrecognition,constructKG}, \cite{ vos2022towards, deem}},
    {group4}/{CAAFE~\cite{Hollmann2023LargeLM}},
    {group5}/{LIDA~\cite{Dibia2023LIDAAT}, InsightPlot~\cite{ma2023insightpilot}},
    {group6}/{ReactionParser~\cite{ko2024filling}, GILT~\cite{nam2024using}, \\ BIRD~\cite{li2024can}  \cite{dubiel2024device, zheng2024revolutionizing}},
    {group7}/{R2E~\cite{jain2024r2e}, PRISE~\cite{zhengprise}, Repoformer~\cite{repocomp}  \cite{KoziolekGHALE24, yuan2023power}},
    {group8}/{Aug-Tree/Linear~\cite{singh2023augmenting} \cite{nam2024optimized}},
    {group9}/{LEFT~\cite{HsuMTW23}, InfiAgent~\cite{HuZWCM0WSXZCY0K24},LaTree~\cite{SongY00024} \cite{EoTGE, bai2024transformers, sayed2024gizaml}},
    {group10}/{PALM~\cite{Wang2024TheoremLlamaTG}, CoqPyt~\cite{Carrott2024CoqPytPN}  \cite{Pei2023CanLL, Wang2024TheoremLlamaTG}},
    {group11}/{LILO~\cite{GrandWBOLTA24}},
    {group12}/{GPTuner~\cite{GPTuner}, CHORUS~\cite{CHORUS}, \cite{tabulargeneration, LLMschema}},
    {group13}/{NL2SQL~\cite{readyforNL2SQL,smalllargemodelNL2SQL,text2sqlevaluation}},
    {group15}/{PFM+SMT~\cite{jha2023counterexample, ye2024satlm}}
  }
  
  \def\yShift{-1cm} 
  
  \foreach \name/\text in \middleNodes {
    \node[middleNode] (\name) at ($(middleStart)-(0,\yShift+\leftVerticalSpacing)$) {\text};
    \pgfmathparse{\yShift+\middleVerticalSpacing}
    \xdef\yShift{\pgfmathresult}
  }
  
  \draw[->, thin, draw=DataManagementColor] (data_management.east) to [out=0, in=180] (group1.west);
  \draw[->, thin, draw=DSLUnderstandingColor] (group1.east) to [out=0, in=180] (dsl_understanding.west);
  
  \draw[->, thin, draw=DataManagementColor] (data_management.east) to [out=0, in=180] (group2.west);
  \draw[->, thin, draw=AssessmentColor] (assessing_results.east) to [out=0, in=180] (group2.west);
  \draw[->, thin, draw=DSLUnderstandingColor] (group2.east) to [out=0, in=180] (dsl_understanding.west);
  
  \draw[->, thin, draw=DataManagementColor] (data_management.east) to [out=0, in=180] (group3.west);
  \draw[->, thin, draw=DataQualityOptimizationColor] (group3.east) to [out=0, in=180] (PFM_quality.west);
  \draw[->, thin, draw=DataQualityOptimizationColor] (group4.east) to [out=0, in=180] (PFM_quality.west);
  
  \draw[->, thin, draw=DataManagementColor] (data_management.east) to [out=0, in=180] (group4.west);
  \draw[->, thin, draw=AutoMLColor] (group4.east) to [out=0, in=180] (automated_ml.west);
  
  \draw[->, thin, draw=ExploratoryDataAnalysisColor] (exploratory_data_analysis.east) to [out=0, in=180] (group5.west);
  \draw[->, thin, draw=DSLUnderstandingColor] (group5.east) to [out=0, in=180] (dsl_understanding.west);
  
  \draw[->, thin, draw=ExploratoryDataAnalysisColor] (exploratory_data_analysis.east) to [out=0, in=180] (group6.west);
  \draw[->, thin, draw=AccessibilityColor] (group6.east) to [out=0, in=180] (accessible_models.west);
  
  \draw[->, thin, draw=ImplementationColor] (implementing_methods.east) to [out=0, in=180] (group7.west);
  \draw[->, thin, draw=DSLUnderstandingColor] (group7.east) to [out=0, in=180] (dsl_understanding.west);
  
  \draw[->, thin, draw=ImplementationColor] (implementing_methods.east) to [out=0, in=180] (group8.west);
  \draw[->, thin, draw=AccessibilityColor] (group8.east) to [out=0, in=180] (accessible_models.west);
  
  \draw[->, thin, draw=ImplementationColor] (implementing_methods.east) to [out=0, in=180] (group9.west);
  \draw[->, thin, draw=AutoMLColor] (group9.east) to [out=0, in=180] (automated_ml.west);
  
  \draw[->, thin, draw=AssessmentColor] (assessing_results.east) to [out=0, in=180] (group10.west);
  \draw[->, thin, draw=DSLUnderstandingColor] (group10.east) to [out=0, in=180] (dsl_understanding.west);
  
  \draw[->, thin, draw=ExploratoryDataAnalysisColor] (exploratory_data_analysis.east) to [out=0, in=180] (group11.west);
  \draw[->, thin, draw=ImplementationColor] (implementing_methods.east) to [out=0, in=180] (group11.west);
  \draw[->, thin, draw=DSLUnderstandingColor] (group11.east) to [out=0, in=180] (dsl_understanding.west);
  \draw[->, thin, draw=AccessibilityColor] (group11.east) to [out=0, in=180] (accessible_models.west);
  
  \draw[->, thin, draw=DataManagementColor] (data_management.east) to [out=0, in=180] (group12.west);
  \draw[->, thin, draw=AccessibilityColor] (group12.east) to [out=0, in=180] (accessible_models.west);
  
  \draw[->, thin, draw=ExploratoryDataAnalysisColor] (exploratory_data_analysis.east) to [out=0, in=180] (group13.west);
  \draw[->, thin, draw=AccessibilityColor] (group13.east) to [out=0, in=180] (accessible_models.west);
  
  \draw[->, thin, draw=AssessmentColor] (assessing_results.east) to [out=0, in=180] (group15.west);
  \draw[->, thin, draw=DSLUnderstandingColor] (group15.east) to [out=0, in=180] (dsl_understanding.west);
  
  \end{tikzpicture}
  \caption{\textbf{An Overview of How PFMs Empower Data Analysis Tasks Through Optimizations.} The figure illustrates the relationships between key data analysis tasks (left column)—Data Management, Exploratory Data Analysis, Implementation, and Assessment—and the optimizations enabled by Pre-trained Foundation Models (PFMs) (right column). The middle column lists representative methods and studies that bridge specific tasks to the corresponding optimizations. Arrows indicate how PFMs address challenges within each task, facilitating the transition towards optimized data analysis processes. This visual framework underscores the multifaceted role of PFMs in systematically enhancing data analysis by connecting practical tasks with advanced optimization strategies and the core role of scaling and consolidating domain-specific language (DSL) during data analysis.}

  \end{figure*}

  \subsection{PFM-based Methods for Data Preparation}\label{sec:data_management}
  
  In \textit{Data Management: A Gentle Introduction—Balancing Theory and Practice}~\cite{van2020data}, the authors present the definition of data management~\cite{international2017dama}.
  
  \paragraph{Definition.} Data management is the development, execution, and supervision of plans, policies, programs, and practices that deliver, control, protect, and enhance the value of data and information assets throughout their lifecycle.
  
  To fit in the skeleton of the topic of data analysis in \S~\ref{sec:skeleton}, we focus on the following aspects of data management: data collection, cleaning, and transformation. Those are methodologies to produce well-prepared data.
  
  Preparing real-world data for data analysis encounters complex problems composed of heterogeneity, semantic and random noise, irregularity, and contradictions. Existing work\cite{fernandez2023large} proposed several strengths and challenges in applying large language models to data management. In this section, we organize surveys for more utilitarian methods and interaction between the management of data and other components, as well as optimizations of data analysis.

  Well-managed data should share several of the same good properties. All these properties are presented as goals of those data management tasks~\cite{wang1996beyond}. Furthermore, the performance of data analysis always highly depends on (1) accuracy, which refers to accurately reflecting the phenomena, (2) completeness, which refers to the collection and storage of all necessary data, this ensures thoroughness of knowledge that can be induced from data, (3) consistency, which refers to that there should not be contradictions between data sources or data collected at different times, (4) identity of the same entity, which refers to entities that share the same description in multiple data sources~\cite{khoshafian1986object} and forms, this involves the unification of multi-modal data, and (5) timeliness, which refers to data not being outdated to reflect current circumstances. 
  
  Data collection aims to provide complete and consistent datasets given a well-defined analytical query. In the field of data manipulation, collection, and discovery, PFMs show powerful potential in merging multi-domain heterogeneity data and knowledge, consisting of domain-specific terminologies, libraries~\cite{Ellis2020DreamCoderGG}, facts~\cite{Tang2024WorldCoderAM}, distributions~\cite{sordoni2024joint}, and external knowledge. In a broader sense, data analysis is a kind of computation from specified input and output states, which should be properly set by data collection toward predefined goals~\cite{turing1936computable}.

  An important way to achieve the goal of data preparation on datasets with quality issues is data cleaning and data wrangling. To enhance data quality by filling in missing values, correcting errors, and decreasing duplications and inconsistencies, IterClean~\cite{ni2024iterclean} propose a PFMs-based method for iteratively data issue detection and reparation that achieves state-of-the-art performance. PFM-based data wrangling solves multiple tasks such as data cleaning, transformation, and integration and serves as an important data management method. \cite{vos2022towards} proposes a low-resource fine-tuning method for task-specific or data-specific data wrangling methods, aiming to reduce high expertise, manual efforts, and storage costs for finetuning. DEEM~\cite{deem} uses PFMs to understand declarative wrangling demands, routing to data of interest and operating with them using generated codes. \cite{qi2024cleanagent} propose an agent for data cleaning, CleanAgent. \cite{FMwrangling} evaluates multiple data wrangling tasks such as missing data imputation and error detection. It can compose DSL for error-type-specific data cleaning to achieve automation for data repair.
  
  PFMs can induce statistically and deduce logical relations~\cite{Wang2023SoloDD, CHORUS, constructKG,KGobjectrecognition,tabulargeneration, entityresolution}, in which records of different real-world entities are sufficiently identified. To enhance the identity of different data sources, \cite{LLMschema} evaluates PFM's capability to match schemas with similar semantics and different column identifiers. ALT-GEN~\cite{tabulargeneration} benchmarks PFMs' capability to search for possible unions of tabular datasets.
  
  The joinability discovery of tables helps create records that contain more attributes. It is helpful for multiple downstream takes, such as data wrangling~\cite{FMwrangling} and active learning~\cite{ActiveLearning}. Benefiting from FPMs, the joinability of tabular data can be discovered in two styles. \textit{Transductive} methods, such as Ditto~\cite{Ditto}, Ember~\cite{Ember}, DeepJoin~\cite{DeepJoin} and Starmie~\cite{Starmie} which directly drive similarity metrics between recodes and columns from fine-tuned PFMs. \cite{Nobari2023DTTAE} proposes DTT, which is also a transductive PFM-based method for finding linkage implied in multi-table datasets, which transforms string to evaluate distance according to direct match with simply few-shot prompting. \textit{Inductive} methods~\cite{jain2022jigsaw, lyu2024automatic} induce codes or rules for implementing identifiers of joinable linkage. \cite{zuo2022spine} proposes another method for finding linkage, and though it is not a PFM-based method, it implies that the program by-example (PBE) style method will be well discussed in future research. 
  
  \cite{constructKG} and \cite{KGobjectrecognition} consolidate reusable patterns reasoning processes mined by PFMs into knowledge graphs for further usage. \cite{Li2024IsPB} shows that PFMs can do PBE by in-context learning. It can be used for tasks such as regular expression construction, string transformation, symbol regression, and formal proof, which points to a new direction in the field of data integration. These discovered relations can be aligned with timestamps from multiple data sources. This can be helpful in inferring timeliness from causal relations or the discovery of timely data from data sources~\cite{Wang2023SoloDD}.
  
  Feature engineering is another important task for helping applications to identify entities and concepts, specific form and representation of data can greatly affect the efficiency of downstream analysis, which is discussed in detail in \S~\ref{sec:auto_ml}. PFMs also help to build data processing pipelines that perform feature engineering. For example, \cite{Hollmann2023LargeLM} proposes a context-aware method that builds automated feature engineering pipelines — CAAFE — which can automatically keep changes according to performance improvement.

  Apart from the quality and effectiveness of the data preparation process, efficiency and data management are other important metrics of data analysis. PFMs can also leverage external knowledge, for tuning data manipulation knobs to boost efficiency. GPTuner~\cite{GPTuner} embeds expert knowledge in the database manual into Bayesian optimization methods to tune knobs that are efficiency sensitive.
  
  These methods show the progress of benefitting data management from PFMs. Meanwhile, this research area is super young and full of challenges for more advanced tools based on PFMs.
  
  \subsection{Exploratory Data Analysis}\label{sec:explore}
  
  Exploratory data analysis (EDA) and data visualization are essential to intuitively understanding the features and patterns of datasets. Exploratory data analysis aims to help analysts retrieve and aggregate patterns of interest interactively, finding suitable abstractions and partitions that are not only stable but also understandable. Several issues and steps are proposed in \textit{Making sense of data—a practical guide to exploratory data analysis and data mining}~\cite{myatt2007making} to be considered to be taken and interactive with each other in any exploratory data analysis: problem definition, data preparation, implementation, deployment. We consider these steps, except for problem definition, to be operated in other tasks. This section focuses on the interaction between analysts and platforms to perform these tasks toward an analytical need and gradual derivations of concise problem definitions.

Problems should be first defined as concisely as possible before the final implementation of any data analysis tasks~\cite{myatt2007making}. This involves constant criticism of the content and form of the problem and the object associated with it until the most rigorous composition is formed, to avoid inconsistencies and ambiguities. For the prior objects involved in data analysis, e.g., directly defined goals, stakeholders, etc., PFMs help to link them together with other objects that should be identified posteriorly after inductions of the datasets. DTT~\cite{Nobari2023DTTAE} and Solo~\cite{Wang2023SoloDD}, which search functional links for column names in relational databases, can also be used for organizing objectives defined to construct a complete, detailed, and concise data analysis problem. For the posterior objects, PFMs help to induce them and deduce the most important ones and their proper names. \cite{ZhangS24} proposes EDC, which induces possible objects grounded in the dataset, then defines them using proper names, and finally canonicalizes them to the most concise forms. Data discovery systems can be built with PFMs such as Solo~\cite{Wang2023SoloDD} and CHORUS~\cite{CHORUS} for finding stable and utilitarian patterns for modeling and retrieval.
  
  EDA helps analysts to better dig into unknown structures (e.g., analytical, geometric, and algebraic properties) implied in complex datasets. For example, modern databases provide multiple analytical interfaces for analysts to help them flexibly query instances. According to the values, by using statistical, machine learning, and graphic methods, EDA helps to identify main features, anomalies, and implicit relations in data. Implementing methods composed of methods routing to valuable parts of data and exhibition methods is filling the gap between general-purpose libraries and ad-hoc exploration solutions which can present a huge cost of manual labor. Declarative methods help to mitigate this obstacle, but real-world data would be so complicated for higher-level abstractions~\cite{heer2010declarative, shih2018declarative, kim2022cicero}. 
  
  There are PFM-based methods that enhance interfaces for EDA, and some of the other methods fill the gap between general-purpose methods and task-specific demands, they all made significant successes. ReAcTable~\cite{ReActTable} enhances tabular data question answering with PFMs. \cite{zheng2024revolutionizing} introduces DQA, a benchmark for evaluating the question-answering performance of PFMs based on databases, which provides a more abstract declarative interface for EDA. Analysts can declare one-off EDA tasks, and then PFMs can infer instances dependent on libraries that can solve sub-tasks and composite the implementation of the task. In terms of data visualization, \cite{ma2023insightpilot} proposes InsightPlot. It can construct different types of insights such as objects, types, and attributions, and allows analysts to gain more insight into complex structures with iterative interaction with PFMs. Combining the ability to do interactive parsing, data routing, and visualization, the introduction of PFMs provides an easy-to-use interface for exploring scalable databases. 
  
  PFMs can help analysts to understand and manipulate data or task-specific operations and data processing pipelines, to explore more complex analytical tools and insights with more complex logical structures. For parsing and routing to valuable parts of datasets, not only can the analysis of tabular datasets benefit from PFMs, but unstructured and semi-structured data can also be more easily explored with the help of PFMs. For example, \cite{ko2024filling} proposes a PFMs-based method for parsing semi-structured data like HTML data into knowledge that is easy to explore. \cite{Dibia2023LIDAAT} proposes LIDA, pointing out that PFMs not only can generate charts by manipulating graphical languages but also provide explanations to novices about the content of images and their insights, offering a human-centered data visualization interface. \cite{nam2024using} introduces a large-model-based code explanation tool GILT, which helps users quickly understand and implement large sections of data analysis code. \cite{GrandWBOLTA24} proposes LILO, which generates doc-strings for obscure imperative codes. This can enhance the feasibility of non-computer experts in understanding and exploring code repositories. NL2SQL~\cite{smalllargemodelNL2SQL,text2sqlevaluation,readyforNL2SQL}, which enhances the accessibility of non-expert to relational databases, can be effectively and efficiently implemented by PFM-based methods. 
  
  PFMs can also help to predict user intentions. This could help to resolve vague and ambiguous analytical intent. \cite{dubiel2024device} propose a query intent prediction method, which can improve the efficiency in contrast with the dump query method without perception of analytical needs. \cite{li2024can} also discussed the feasibility of using large language models as database interfaces. Similar to which involves encoding human intentions represented by natural language into query statements. These methods perform abductive reasoning to provide the most useful query advice according to the present goals of data analysis.
  
  Exploratory data analysis is needed in various industries, covering both structured and semi-structured data. PFMs can parse semi-structured data through direct or encoded methods, while structured data is more suited for exploration and analysis through rule-based approaches. In \S~\ref{sec:interface}, we put stress on the potential for systematically enhancing interfaces for data analysis which can promote the development of EDA. In \S~\ref{sec:interface}, we generalize the boundary of exploration for not only the dataset but also concepts that can be learned from data and delve into interpretability and editability, demonstrating how PFMs enhance rule-based or principle-based machine learning and its advantages.
  
  \subsection{Implementing Data Analysis Methods and Models}\label{sec:da_methods}
  
  In contrast to EDA where analytical objects are always unknown or vague at first and need to be inferred and unambiguously fixed, data analysis methods include widely used statistical methods and machine learning methods, composed of needed knowledge and mathematical structures. Complicated veins make it hard to systematically and universally represent and fully use these methods. Choosing a suitable one depends on the problem properties and the data characteristics. For example, linear regression is designed for predicting continuous data, while decision trees are designed for classification~\cite{032hastie2009elements}. This selection of models can be seen as choosing different hypothesis classes for approximately identifying concepts of interest, which is detailed in \S~\ref{sec:interface}. 
  
  Choosing methods and models for well-defined tasks is an optimization over a disjoint union of multiple model classes and their parameters. The second concern with the models is the explainability and editability, which would be essential for high responsibility and efficiency. External logical knowledge is also hard to obtain without domain expertise~\cite{031guyon2003introduction}. Learned parameters and rules can be difficult to understand, meanwhile, it's hard to effectively and efficiently inject external knowledge for better performance or bias reduction, and transfer learned concepts for further analysis. The lack of interpretability presents practical and ethical challenges to data analysis.

  To solve these challenges presented above, PFMs introduce formal reasoning-based methods, which show some advances in understanding analysis methods and data to be analyzed. PFMs can help to search hyperparameters to guide machine learning with higher performance~\cite{sayed2024gizaml}. Furthermore, PFMs can also perform as statisticians who perform in-context algorithm selection with provable performance~\cite{bai2024transformers} by mimicking gradient descent during inference. \cite{HuZWCM0WSXZCY0K24} proposes a benchmark — InfiAgent — and an agent for data analysis — DAAgent — which involves multiple analytical tasks such as outlier detection, distribution analysis, and machine learning, etc. \cite{EoTGE} propose guided evolution (GE), which modifies codes for PyTorch models using PFM directly to evolve the models. 
  
  Consolidating the inference of PFMs into simpler methods and models helps to gain interpretability. \cite{singh2023augmenting} proposes Aug-Linear and Aug-Tree which make predictions based on N-grams in text. Predicted responses of voxel (specific points of the human brain where the strength of activity is to be measured) have exceeded BERT performance with simple linear models. \cite{nam2024optimized} proposes OCTree, which augments tabular data for improving downstream performance. 
  
  Domain-specific data analysis methods often have stringent requirements~\cite{020arlot2010survey}, while general analysis methods lack the formalization and discussion of external knowledge. Effective analysis requires transcending statistical generalization methods and gains the support of mathematical logic's formal validity and ambiguity. \cite{HsuMTW23} proposed LEFT, which consolidates concepts in some domain (2D movement) to general concepts across domains (3D movement). \cite{SongY00024} proposed a tree-based variational inference method to encode event sequences into strict logic expressions. \cite{WangCY0L0J24} proposes CodeAct, which substitutes PFM actions into executable codes, producing a more rigorous decision implementation. \cite{GrandWBOLTA24} introduces LILO, which produces docstrings for complex implementations that can improve interpretability and help PFMs reasoning by inducing knowledge directly with observations from these docstrings. These docstrings are representations of implementations for a group of thoroughly solved problems~\cite{yuan2023power}. Their composed structures represent possible analysis solutions~\cite{zhengprise}.
  
  More generally, code generation makes it possible to deduce principles from grounded intentions of data analysis directly and consolidate them with programming language~\cite{KoziolekGHALE24}. The robustness and stability of traditional data analysis models are key to ensuring their reliable application in different environments. From a more systematic view, PFMs can consolidate knowledge more imperatively through code repositories. PFMs gain more and more attention in understanding and repositories. \cite{jain2024r2e} proposes R2E which turns repositories into PFMs agent environments. \cite{repocomp} proposes Repoformer, which can solve code completion in the user's repository.

  \subsection{Assessing Analyzed Results} \label{sec:correctness}
  
  Correctness and applicability are the most important criteria for data analysis, reflecting the reliability of data analysis results. In this section, we refer to correctness as implementations, statements, proofs, and derived theories that work correctly without inconsistencies and exceptions computationally and logically, according to axioms and principles. The applicability of the results determines whether they can effectively guide decisions and actions \cite{027davenport2017competing}, which are translated into concrete action plans \cite{040shmueli2011predictive}.
  
  In real-world problems, bridging the gap between data analysis and practical applications still requires substantial labor and complex manual methods. It involves validity tests, controlling termination, verification, and proof. Solving real-world problems often involves complex nested and recursive logic, and finding analysis methods that are globally, locally, and necessarily correct \cite{shao2023synthetic} is the main challenge in handling these tasks \cite{yao2024tree}.

  \begin{figure*}[h]
    \centering
    \includegraphics[width=0.8\textwidth]{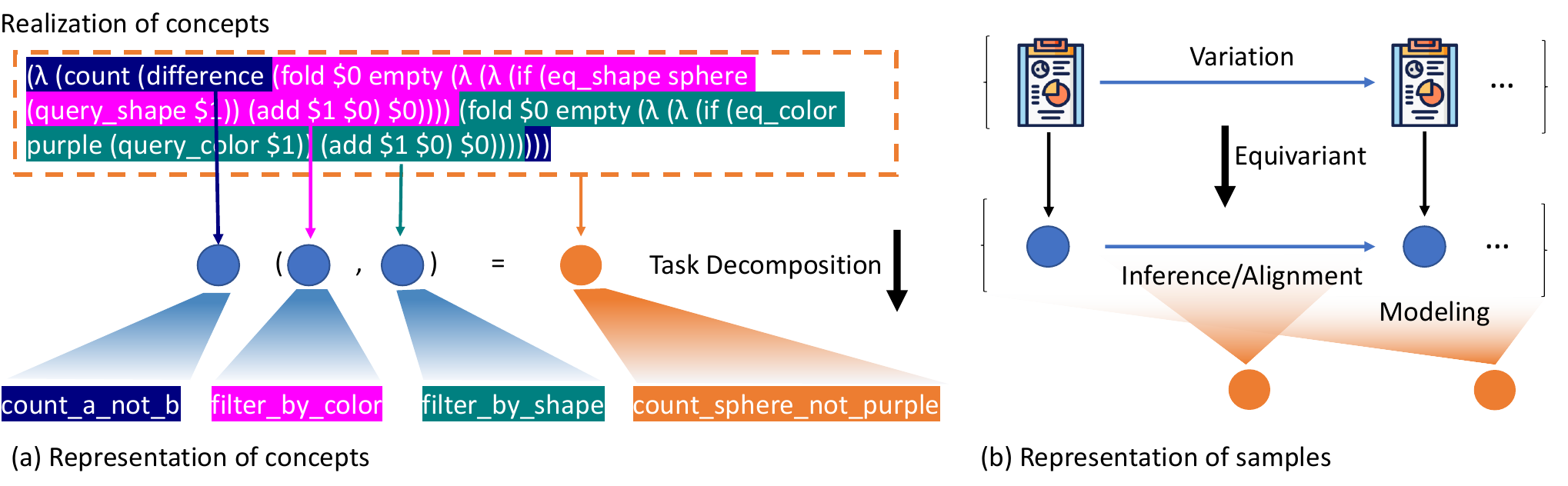} 
    \caption{\textbf{Representation learning of concepts and samples.} (a) Implementation of concepts in one computational model can be represented by more concise symbolic forms. (e.g. func-names in compiled libraries.) (b) Structures of interest can be extracted and aligned as inference and modeling processes. They are identified by  Representation learning is recompiling and compressing algorithms and datasets.}
    \label{fig:representation_learning}
  \end{figure*}
  
  Accessing results from data analysis involves statistical tests for quantized results or unit tests for bug detection~\cite{schafer2023empirical,alshahwan2024automated}. They involve the partitioning of equivalent classes of datasets or discuss the influence of outliers. Some of them involve mathematical proofs that provide assessments of thoroughness for a group of solving problems, which can always be seen in software tests~\cite{schumann2013automated}. PFMs-based code generation and theorem-proving systems \cite{Pei2023CanLL,Wang2024TheoremLlamaTG, Rodriguez2024ExploringAO} make it possible to develop and assess new self-organizing data analysis methods. 
  
  PFMs can not only be combined with executable languages but can also benefit from satisfiability modulo theories (SMT)~\cite{ye2024satlm}. Adversarial examples~\cite{jha2023counterexample} can be searched within the scope of data analysis but can not be solved properly by the current solution. These exceptions hurt the completeness and robustness of data analysis results. Methods that introduce satisfiability theories are first introduced data analysis and engineering with high demand and risk, e.g., financial risk analysis, medical diagnostic and analysis, etc. For non-expert users with higher ambition, PFMs help to generate formal representations that can be identified by SMT solvers to gain consistency, robustness, and fairness during the assessment and further optimization.

  Data analysis is a process of continuous improvement and iteration. By establishing feedback mechanisms, optimization directions can be constantly gathered from practical applications, allowing for model optimization and adjustment. Iterative optimization emphasizes continuous improvement in aspects such as model performance, data quality, and analysis methods to adapt to changing environments and needs \cite{033kaelbling1996reinforcement}.

  This places greater demands on quickly obtaining feedback according to assessment results from the environment. The cross-domain integration driven by PFMs not only enhances the comprehensiveness and depth of data analysis but also facilitates communication and collaboration among professionals from different backgrounds by providing interpretable analysis results. For example, PFMs can be used to integrate data from different fields to form a comprehensive analysis framework, thereby offering more comprehensive and accurate insights~\cite{yang2024give, hu2023survey}.

  \section{PFMs Enhanced Systematic Optimization Methodologies}~\label{sec:methods}
  
  
  In this chapter, we will explore how pre-trained foundation models systematically optimize data analysis. From reasoning, accessibility of complex structural data and models, and data quality optimization, to automated machine learning, PLMs have demonstrated significant potential at every stage. Nevertheless, we also point out the current challenges and emphasize the importance of future research in algorithms, data quality, and system efficiency. Through these studies, we hope to drive data analysis systems toward becoming more efficient and robust.
  
  \subsection{PFM Enhance Reasoning}\label{sec:dsl}
  
  "The basis of all human culture is language, and mathematics is a special kind of linguistic activity." Arnold \& Manin (2000)
  
Domain-specific language (DSL) refers to languages specifically designed for a specific application domain. It reflects specialization, abstraction, and effective scope of human understanding, improving problem-solving and communication. It narrows down the reasoning process into the most important and meaningful proposition space. Vapnik proves that a machine has only two mechanisms to learn, which are minimizing empirical loss and minimizing confidence interval~\cite{vapnik2019rethinking}. The latter commutes external predicative prior knowledge into learning, with specific domain abstraction and relational knowledge. For every data analysis expert in any domain, reasoning about reality is a fundamental task, which plays an important role in every stage of data analysis. Till now, in domains like supply chain, legal, and financial field, it costs multiple years of training and experience accumulation for the experts to understand the objects and operations in a domain-specific language fully~\cite{garcia2010using}. 
  
In this section, we consider concepts about data analysis that scales with reasoning. First, we introduce representation learning techniques that identify and encode concepts and entities with symbols and geometric signatures, i.e., vectors~\cite{TURL} or tensors, (\S~\ref{sec:representation}). Then we introduce categories of reasoning, which provide general tools for inference, planning, and code generating (\S~\ref{sec:reasoning}). Finally, we discuss the management of solutions of methods with different structural properties with the evolution of a system and growing analysis demands by consolidation (\S~\ref{sec:consolidate_dsl}).

\subsubsection{Representation learning of concepts.} \label{sec:representation}
  
\begin{figure*}[h]
    \centering
    \includegraphics[width=0.65\textwidth]{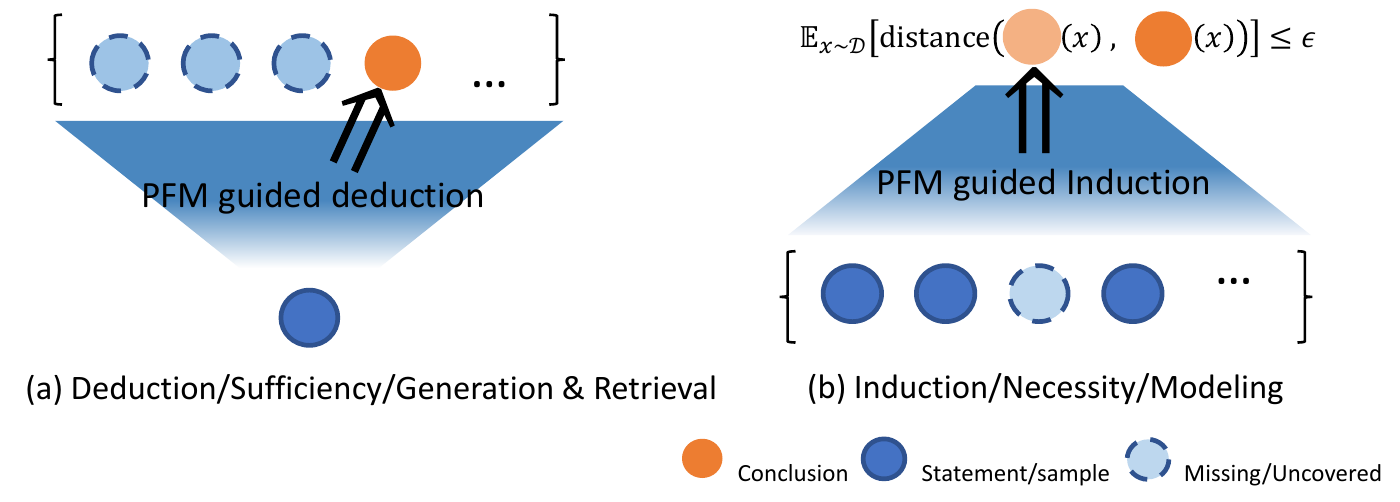} 
    \caption{\textbf{Two kinds of reasoning.} (a) Deduction provides top-down reasoning that proves sufficiency between statements and conclusions, which generates special cases/samples according to universal principles and rules or hypotheses. (b) Induction provides bottom-up reasoning that proves the necessity between samples and conclusions, which approximate and conclude principles and rules from cases/samples. Other kinds of reasoning can be seen as compositions of these two kinds of reasoning while introducing inconsistencies and approximations to compromise consistency and completeness.}
    \label{fig:logic}
\end{figure*}
  
Representation learning in a narrow sense refers to techniques like manifold learning which preserves the geometric shape of the original distribution in support of classification, regression, clustering, and data generation.~\cite{buchholzrobustness, feyposition} In a general sense, representation learning can be related to representation theories in the language of mathematics, which preserves morphisms and structures of categories, preserving symmetry and equivariance, providing algebraic tools, and enhancing optimization and presentation abilities~\cite{yuan2023power, blaauwbroekgraph2tac, HansenCF24}. 
  
Representation learning provides us with tools to compress data and implement algorithms in a concise form, which can be seen as a new DSL. In terms of data compression, \cite{DeletangRDCGMGW24} has proven that language models can be seen as general compression algorithms. It shows that language models can perform lossless compression on images, audio, and videos, which can be far more efficient than conventional algorithms. \cite{GrandWBOLTA24} propose a code compressing and documenting method, LILO, based on PFMs, which can consolidate code into more reusable and readable forms. \cite{zhengprise} propose PRISE which is a compression algorithm for temporal actions made by agents, abstracting valid components into skills that can be seen as representing algorithms discovered from data. Using this, analysts can perform efficient and precise retrieval of both data and composable operations on them, \cite{KoziolekGHALE24} propose a retrieval argumentative method for control code of automated systems which bound to representations that supply to domain-specific standards. Numeric or geometric representations can be precisely leveraged according to different usage. For example, estimated cardinality can be seen as signatures for devices to distribute computational resources. \cite{FMcard} proves that PFMs can also be leveraged for the estimation of cardinality.
  
The introduction of representable rules and principles makes it easy to leverage what is discovered from one set of data to analyze others~\cite{yuan2023power}. Analysts can easily transduce this explicit knowledge on demand. Causes and their effects can be found in the dataset leveraging causal or interventional representation learning in an interpretable and grounded way~\cite{ahuja2023interventional}. They negate existing goals of the analysis to create new goals that are more feasible. To systematically boost the ability of data analysis on more complex systems like code repositories~\cite{jain2024r2e}, disciplinary knowledge of specific domains, and their application fields, we need more foundational representation theories~\cite{yuan2023power} that PFMs can first bring in.

\subsubsection{Reasoning with domain-specific concepts}\label{sec:reasoning}

Representation learning for DSL contributes significantly to the efficiency and quality of analysis and decision-making, emphasizing both inductive and deductive logic. Reasoning involves composing these representations toward a predefined goal, which not only induces evidence from datasets and adjusts the posterior but also deduces applicable rules to reach necessarily correct conclusions. 
  
This section summarizes how PFMs apply to principles and conditions of induction and deduction in reasoning, making systematic enhancements in the field of data analysis. We introduce inductive reasoning which provides cogency from necessarily satisfied cases to general rules, and deductive reasoning which sufficiently produces special cases from general principles and evaluates the validity of the existing statements.
  
\paragraph{Principles of Induction.} Inductive reasoning involves deriving general principles from specific observations or instances of collected data. In finite or infinite deterministic systems, such as mathematical objects or physical laws, the inductive method can identify deterministic principles and rules with certainty. However, in stochastic systems — such as datasets with uncertainties, as discussed in \S~\ref{sec:data_quality} — we can only pursue hypotheses that are approximately correct. These hypotheses are subject to validation and may need to be revised as more data becomes available, which will be detailed in \S~\ref{sec:interpretability}.

In real-world applications, successful inductive reasoning relies on several key principles:
  
\underline{Completeness of representative sampling.} Ensuring that the data collected accurately and sufficiently reflects the population or phenomenon being studied. This reduces sampling bias and enhances the generalizability of the inductive conclusions~\cite{037little2019statistical}. PMFs can help to perform completeness checking by encoding an induction procedure into a satisfiability problem~\cite{ye2024satlm} which can be solved by SMT solvers and finally guide the inductive steps with abductive style counterexamples~\cite{jha2023counterexample}. In a real-world scenario, full completeness can be not achievable, where all possible samples are discussed or covered by equivalent classes discussed. These inductions don't provide sufficiency for precisely correct results, which introduces statistics to produce PAC learning.
      
\underline{Consistency and robustness against errors.} High-quality data and resilience to errors are essential for reliable inductive reasoning. PFMs contribute by automating data cleaning and preprocessing tasks, such as handling missing values, correcting errors, and removing inconsistencies~\cite{ni2024iterclean, qi2024cleanagent}. This improves the accuracy of the induced rules and principles.
      
\underline{Compatible logical structures and sufficient prior} \\ \underline{knowledge.} Structures of the space where principles are to be searched should not be too small to contain the precise form. It should not be too large to identify these concepts with a limited number of cases. PFMs assist in identifying hidden patterns and relationships that might be overlooked using traditional methods~\cite{ma2023insightpilot, Dibia2023LIDAAT}. These logical structures can be leveraged to integrate existing domain knowledge to inform the inductive reasoning process. PFMs can merge multi-domain heterogeneous data and knowledge bases, enhancing the richness of the inductive models~\cite{Ellis2020DreamCoderGG, Tang2024WorldCoderAM}. This leads to more guided and consistent induction steps.
      
\underline{Well-defined goals.} Having clear inquiries with rigor synthesis, which is a proposition formed by integrating multiple possible choices, e.g., a thesis and its antithesis, for the inductive analysis ensures that the reasoning process is focused and relevant. PFMs can interpret natural language descriptions of goals and translate them into actionable analytical tasks~\cite{li2023resdsql, gu2023few}. Moreover, ensuring that hypotheses are falsifiable is essential for scientific rigor.
  
By adhering to these principles, inductive reasoning becomes more robust and reliable. PFMs enhance this process by providing advanced tools for data management, pattern recognition, and knowledge integration. They effectively handle complexities such as heterogeneity, noise, and inconsistencies in real-world data~\cite{vos2022towards, deem}, thereby improving the completeness and consistency essential for sound inductive reasoning. This empowers analysts to derive meaningful insights and develop approximately correct hypotheses, even in the face of data imperfections and uncertainties.

  \begin{figure*}[h]
    \centering
    \includegraphics[width=0.8\textwidth]{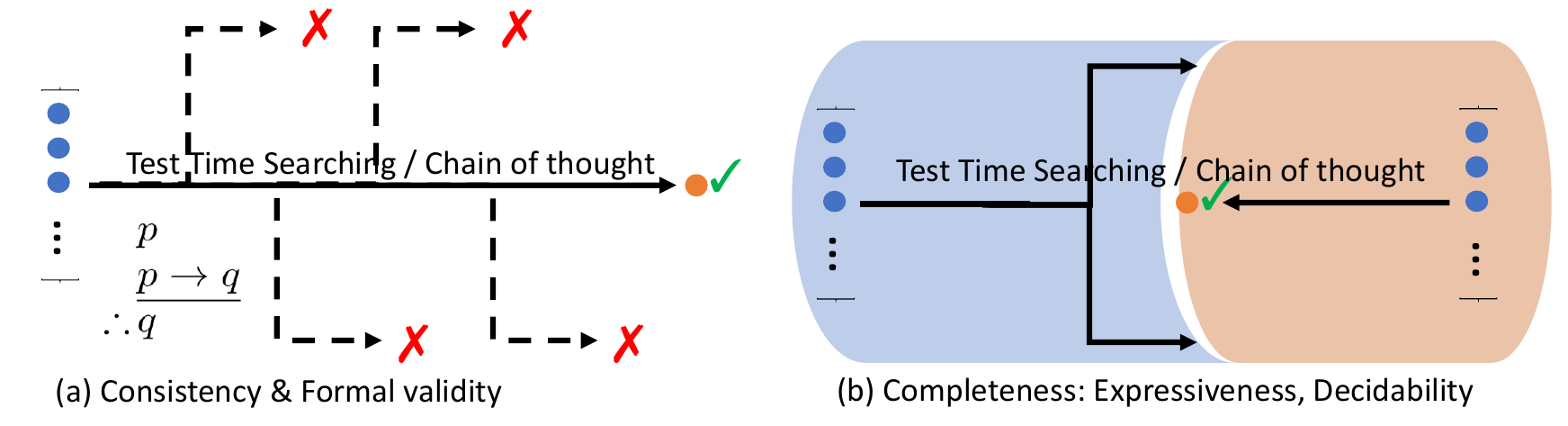} 
    \caption{\textbf{PFM-based reasoning algorithm.} (a) True statements can be produced by inconsistent reasoning due to high validity. E.g., $q$ is necessarily satisfied according to $p, p\rightarrow q$, which provides formal validity from classical logic. (b) Adjusting the expressiveness by approximation compromises the decidability and completeness of the reasoning algorithm. Essential factors for PFMs' augmented reasoning lie in these mechanisms.}
    \label{fig:logic_system}
  \end{figure*}

  \paragraph{Principles of deduction.} Deductive reasoning converts principles and axioms, i.e., symmetry or equivariance induced from data, into specific rules that reason about specific instances. Deductive reasoning provides the foundation for deriving necessarily correct conclusions from established principles. When applying deductive reasoning, it is necessary and sufficient to consider several fundamental principles. Reasoning based on deduction can be feasible and correct when the axioms are grounded. Local properties of successful deduction include soundness, which ensures the preconditions are correct, and formal validity, which ensures the rules used are concise and unambiguous. Global properties include consistency, completeness, and ensuring that no contradictions exist in propositions, as well as all true statements can be proven using axioms and rules.
  
  \underline{Fulfilling local properties.} PFMs enhance deductive reasoning by assisting in formal reasoning-based methods, which are crucial for understanding analysis methods and the data to be analyzed and locally depend on formal validity and soundness of preconditions. For example, PFMs can generate formal proofs and Turing-complete formal languages and code~\cite{Li2024IsPB} which have high formal validity, improving the efficiency of problem-solving by constructing solutions that conform to nested and recursive logic~\cite{Khakhar2023PACPS}. This allows for more rigorous and complete analysis methods, helping users address a wide range of data quality issues and build bridges between data analysis problems and specific implementations to achieve precise and unambiguous results~\cite{qi2024cleanagent}. Moreover,~\cite {Li2024IsPB} have shown that PFMs can treat each inference as inductive and possibly correct in-context learning process, which also shows the ability for the discovery of concepts that would enhance soundness in real-world applications.
  
  However, challenges arise due to inherent limitations in logical systems, as highlighted by Gödel's incompleteness theorems, which state that any sufficiently expressive formal system cannot be both complete and consistent. This implies that when we enforce strict consistency in our logical frameworks, we may sacrifice completeness—the ability to derive all true statements. In the context of data analysis, imposing such strong constraints of absolute consistency might limit the system's capacity to reason about complex or contradictory data, potentially diminishing the depth and scope of analysis.
  
  \underline{Compromising of consistency.} To address this issue, some analysts introduce paraconsistent logic, a non-classical logic that tolerates contradictions without leading to triviality — where any statement becomes provable. This approach allows reasoning processes to continue effectively even in the presence of inconsistencies, which are common in real-world datasets due to noise, errors, and conflicting information~\cite{022batini2009methodologies}. By incorporating paraconsistent logic, PFMs can handle contradictory data more robustly, enhancing the reliability of analytical outcomes~\cite{037little2019statistical}.
  
  \underline{Compromising of completeness.} Practitioners must therefore balance between expressiveness—the ability of the logical system to represent complex concepts—and decidability—the feasibility of algorithmically determining the truth value of statements within the system. An overly expressive system may become undecidable, making it computationally impractical for analysis. PFMs contribute to achieving this balance by leveraging advanced reasoning capabilities that handle complex, expressive representations while maintaining computational efficiency~\cite{Pei2023CanLL, Wang2024TheoremLlamaTG}. This balance is crucial in designing logical systems for data analysis that are both powerful and practical, enabling more flexible and comprehensive analysis of complex datasets.
  
  Moreover, the robustness and stability of analytical models are essential for their reliable application in different environments~\cite{029dietterich1995overfitting}. PFMs contribute to enhancing the robustness and generalization ability of models by incorporating external logical knowledge, which is often hard to disclose from data distribution and challenging to obtain without domain expertise~\cite{031guyon2003introduction}. By leveraging mathematical derivations and formal proofs~\cite{wang2023large}, PFMs help analytical approaches achieve greater robustness and generalization performance, thereby strengthening deductive reasoning in data analysis.
  
  Reasoning reveals what's inside datasets with completeness and details, they can be composed of complex reasoning procedures like transductive learning. They can be used to analyze and manipulate biases of complex and black box models. PFMs help us to revisit reasoning to generalize and scale the scope of data analysis.
  
  \subsubsection{Consolidation of DSL}\label{sec:consolidate_dsl}
  
  In support of reasoning, PFMs can perform as components of a family of logical systems, in which we can balance incompatible principles of complex systems — consistency and completeness. Representation learning and reasoning can enhance each other toward the goal of analysis. As an important part of organizing partial solutions, In terms of consolidation of DSLs, PFMs can be used as automated administrators of repositories and knowledge bases. Data analysts should always decide which principles and implementations of frequently used operations or sequences of decisions are important and should be preserved. These can be compositions of top-down deductive reasoning of principles and goals and bottom-up inductive reasoning of examples and applications with formal or latent representations. There remain three challenges for consolidating DSLs efficiently and effectively.
  
  \paragraph{Reusability.} Reusability is a cornerstone for efficient DSL consolidation, enabling components and frameworks to be applied across multiple contexts and applications. By promoting modular design and standardization, PFMs can create canonical and adaptable code segments. For instance, PFMs can aid in generating reusable code repositories~\cite{jain2024r2e, repocomp}, and in creating generalizable data analysis methods that can be adapted to different datasets and tasks. The approach to managing consolidated domain knowledge can greatly influence the reusability. Hopefully, PFMs can leverage knowledge graph~\cite{constructKG,KGobjectrecognition} and vector databases~\cite{vectorstorage} to manage complex non-structural knowledge for efficient responsibility.
  
  \paragraph{Understandable.} Making consolidated DSLs understandable is crucial for user adoption and effective implementation. This involves presenting complex logical structures and reasoning processes in a clear and accessible manner. PFMs can contribute by generating human-readable documentation and offering explanations of underlying algorithms and decisions. Additionally, incorporating intuitive interfaces and visualization tools can aid users in comprehending and interacting with the DSL. For example, PFMs can assist in exploratory data analysis by generating insights and visualizations~\cite{ma2023insightpilot, Dibia2023LIDAAT}, helping users to better understand complex data structures.
  
  \paragraph{Thoroughness.} To ensure thoroughness in consolidating DSLs, it is imperative to capture the full scope of domain-specific knowledge and operations without overlooking critical details. For mathematical definitions, thoroughness refers to all circumstances being fully discussed, at least important circumstances are well covered for an approximation~\cite{brand2023parameterized}. This involves a meticulous examination of existing principles, methodologies, and exceptions within the domain. PFMs can assist by systematically analyzing large datasets to identify patterns and gaps in knowledge bases, thereby highlighting areas that require further attention. Additionally, PFMs can help in managing the quality of implemented methods and software by assisting in performing unit tests and mathematical proofs~\cite{Wang2024TheoremLlamaTG, Carrott2024CoqPytPN}, ensuring that the consolidated DSL reflects accurate and comprehensive knowledge. Achieving thoroughness demands a balance between comprehensiveness and practicality to prevent information overload and maintain system efficiency.

  \subsection{PFM empowered Accessibility}\label{sec:interpretability}
  
  \begin{figure*}[h]
    \centering
    \includegraphics[width=0.65\textwidth]{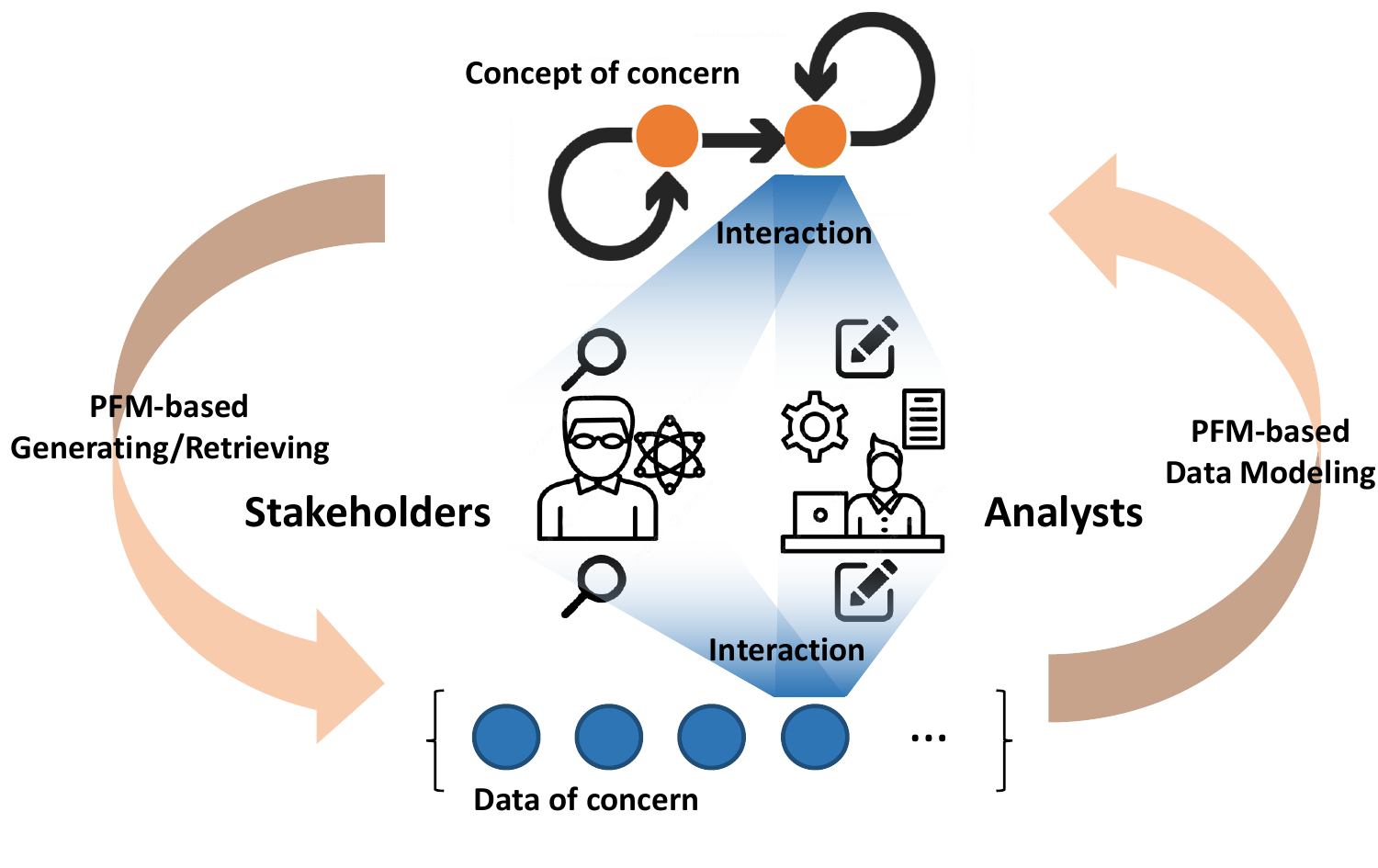} 
    \caption{\textbf{Overview of accessibility.} Successful data analysis involves stakeholders and analysts. Stakeholders' concern about the achievement of the goals proposed. Analysts are responsible for the implementation and completeness of data analysis. Which involves modeling and retrieval/generation to discover universal laws and principles implied in the structure of datasets. Interactions and accessibility should be efficiently and effectively introduced.}
    \label{fig:accessible}
  \end{figure*}
  
  Building upon the foundations of reasoning and the consolidation challenges discussed previously, in this section we discuss the transformative capabilities offered by PFMs for data analysis interfaces. By integrating advanced representation learning and reasoning, PFMs enhance the way analysts interact with data and analytical tools. 
  
  Then, we introduce the concept of probably approximately correct (PAC) from machine learning theory. Philosophically, the PAC theory encapsulates the ontological inquiry into real entities by quantifying the capacity of learning algorithms to approximate true concepts from finite samples. Simultaneously, it emphasizes the process of distilling essence from phenomena. In this section, we discuss the transparency and understandability of the learning procedure (\S~\ref{sec:interface})  and how automatically learned concepts in PAC learned concepts contribute to interpretable and editable machine learning (\S~\ref{sec:editability}).
  
  \subsubsection{PFMs Enhanced Data Analysis Interface}\label{sec:interface}
  
  Enhancing the data analysis interface not only improves data analysis efficiency but also helps users gain deeper insights into data. This includes interfaces for data manipulation, data visualization, data integration, data preprocessing and cleaning, machine learning and modeling, data exploration, real-time analysis and monitoring, and collaboration and sharing. PFMs provide an operational interface that offers significant benefits for human interaction~\cite{dubiel2024device}, serving as a bridge between human preference choices and real-world application problems.
  
  This section explores how PFMs improve data analysis interfaces through intelligent routing to relevant data and methods, automated documentation, translation, and enhanced interaction emphasizing interpretability,  and operability.
  
  \paragraph{Routing to Data of Concern}
  
  Accessing the right data efficiently is a critical step in any analytical process. PFMs can significantly streamline this step by intelligently routing analysts to the data of concern. Leveraging their strong semantic understanding capabilities, PFMs interpret user queries to identify and retrieve relevant datasets from extensive repositories, even when dealing with different structured and semi-structured data types that often have different data manipulation interfaces.
  
  For example, an analyst may inquire, "Find me the latest sales data for the Northeast region excluding returns." The PFM can parse this request, understand the specific requirements, and locate the precise dataset that matches these criteria. This capability reduces the time spent on manual searches and minimizes the risk of overlooking pertinent data, thereby enhancing the thoroughness of the analysis as discussed in the consolidation of DSLs.
  
  PFMs also assist users in interacting with data analysis systems by establishing a mapping between business logic and data manipulation and exploration methods. Different data types require different interfaces; for instance, relational databases use Structured Query Language (SQL) and its dialects like PL/SQL. PFMs excel in generating data queries~\cite{li2023resdsql, gu2023few, cheng2022binding,smalllargemodelNL2SQL,text2sqlevaluation,readyforNL2SQL,CatSQL}, allowing users to formulate complex SQL queries through natural language instructions. This not only makes data retrieval more accessible but also reduces the likelihood of errors in query formulation. Database tuning plays an essential role in advancing efficient access of very large databases. PFMs enhanced methods such as GPTuner~\cite{GPTuner} help to inject external knowledge into such systems to gain efficiency~\cite{NLPTune}.
  
  For semi-structured and unstructured data, languages such as Cypher and Gremlin are used for querying graph databases. PFMs can aid in formulating these queries as well, enhancing the performance of systems that rely on knowledge graphs~\cite{huang2023kosa}. Instead of using hard similarity or word frequency-based methods, information retrieval methods neural networks align patterns in multiple granuarity~\cite{NeuralDB}. Additionally, specialized query languages like SPARQL and XQuery are used to interact with knowledge bases and XML data. By utilizing PFMs, analysts can interact with these systems using more user-friendly natural language commands~\cite{li2024flexkbqa, Lehmann2023LanguageMA,ReActTable}. 

  PFMs' can also leverage semantics that can not be directly captured by human-curated frameworks and algorithms, which are encoded by datasets with complex and ambiguous structures. These datasets are always defined as non-structural data. Schema mapping~\cite{LLMschema} can be seen as tasks performed on datasets composed of incompatible tabular data and composable or similar semantics.
  
  \paragraph{Routing to Methods of Concern}
  
  Identifying appropriate analytical methods is as important as accessing the right data. PFMs assist by recommending methods that align with the specific problem context and data characteristics. By understanding the DSL and the underlying principles of various analytical techniques, PFMs guide analysts toward the most suitable tools and methodologies.
  
  For example, when working with statistical analysis, PFMs can suggest utilizing data frameworks such as Pandas and NumPy~\cite{lai2023ds}, interpreting natural language commands to perform data manipulation. If the task involves full-text search, log analysis, or real-time monitoring, PFMs can recommend the combination of search and analysis engines like Elasticsearch, significantly benefiting information retrieval~\cite{zhu2024retrieval}.
  
  Furthermore, PFMs can serve as an interface between different dialects of systems. They can convert code or queries from one language to another, facilitating interoperability. An example is Mallet~\cite{ngom2024mallet}, a method that acts as a code converter, capable of converting queries or code snippets between different languages, thus aiding analysts who work across multiple systems.
  
  \paragraph{Documenting and Translation}
  
  Documentation is essential for knowledge sharing and maintaining the understandability of analytical processes. PFMs can automate documentation by generating detailed explanations of the steps taken during analysis, the reasoning behind method selection, and interpretations of results. They can also optimize existing interaction processes, participating in the optimization of program compilation and data manipulation processes~\cite{cummins2023large, cummins2024meta, li2024can}, thereby achieving significant computational performance improvements. Auto-documentation system helps to generate human-readable functions and variable names and docstrings which capture global structures of the repository, improving interpretability and improving downstream PFMs-guided search~\cite{GrandWBOLTA24}.
  
  Moreover, PFMs can help analysts correct errors in interface languages. Due to the universal tendency of people to make biased choices, the interface language programs written by people often contain subtle errors. Researchers have used PFMs to correct errors in the interface language written by people, achieving significant results. For example, PFMs can help data analysts modify Excel formulas, quickly identifying the causes of formula failures and undiscovered errors~\cite{Bavishi2022NeurosymbolicRF}.
  
  Additionally, PFMs can translate complex technical jargon into accessible language for non-expert stakeholders. They can also facilitate cross-lingual translation, making the findings available to a global audience. This capability ensures that the insights derived are understandable, addressing one of the key challenges in DSL consolidation.
  
  \paragraph{Interaction}
  
  Enhanced interaction between analysts and analytical tools is crucial for effective data analysis. PFMs contribute to this by providing interfaces that are both interpretable and operable. They offer explanations for their recommendations and decisions, making the analytical process transparent.
  
  PFMs enable analysts to interact with data and models using natural language commands. This lowers the barrier to advanced analytical techniques, allowing analysts to modify parameters, run simulations, and visualize data without deep technical expertise. For instance, in the field of statistics and analysis, PFMs can interpret natural language commands to perform data manipulation using frameworks like Pandas and NumPy~\cite{lai2023ds}. This interactive capability supports the balance between expressiveness and decidability in reasoning processes, as previously discussed.
  
  Researchers also utilize PFMs to optimize existing interaction processes. By determining conservative conditions through formal proof and verification, and by providing continuous system analysis and optimization suggestions, PFMs help analysts gradually approach optimal operations. Participation in the optimization of program compilation leads to significant computational performance improvements~\cite{cummins2023large, cummins2024meta}, and this finding applies to the optimization of data manipulation processes~\cite{li2024can}.
  
  By integrating these enhancements, PFMs transform the data analysis interface into a more intuitive, efficient, and effective environment. They address the challenges of thoroughness, reusability, and understandability in DSL consolidation, ultimately empowering analysts to derive deeper insights and make informed decisions.

  \subsubsection{Interpretable and Editable Method and Models}\label{sec:editability}
  
  Machine learning is an effective tool for automating data analysis. Its definition can encompass the objectives of data analysis. The following will explore the systematic optimizations that PFMs can bring to the field of machine learning, including optimizations in data quality, machine learning, interpretable machine learning, and the enhancement of large models for automated machine learning. Therefore, we will start from the theory of probabilistic approximation, dividing current work and directions into the manipulation of concept classes, hypothesis spaces, data distributions, and learning algorithms.
  
  \paragraph{Definition.} In the basic probabilistic approximation theory in machine learning theory, the most fundamental concept is PAC learning. Given a concept class $\mathcal{C}$, a distribution $\mathcal{D}$, and a hypothesis space $\mathcal{H}$. For $\forall c \in \mathcal{C}$, if there exists a learning algorithm $\mathcal{L}$, whose output hypothesis $h \in \mathcal{H}$ satisfies for $0<\epsilon$, $\delta < 1$ 
  
  \begin{equation} P(\mathbb{E}_{x\sim \mathcal{D}}[\text{distance}(h(x), c(x))] \leq \epsilon) \geq 1-\delta \end{equation}
  
  where distance measures the difference between the hypothesis and the concept.
  
  \paragraph{Insight.} This definition discusses the goal of machine learning, which is to learn approximately correct concept classes from data distributions. A good analysis method also requires good concept classes, data distributions, hypothesis spaces, and learning algorithms. Ideally, the hypothesis space $\mathcal{H}$ exactly covers the concept $c$ that needs to be learned. However, this is often not achievable.

  PFMs have the potential to address these challenges by utilizing human-understandable concept classes to construct hypothesis spaces that better meet analytical needs, thereby enhancing the interpretability and editability of machine-learning models. As discussed in \S~\ref{sec:data_quality}, PFMs can improve data quality by representative sampling, generating and helping to build methods that are robust to certain error types, and in \S~\ref{sec:auto_ml}, we see that PFMs serve as learning algorithms explicitly organize discovered and prompted knowledge from analysts, helping them automatically construct machine learning pipelines.
  
  Black box models and decisions lead to undesirable consequences: data analysts and engineers cannot trace the causes of errors, resulting in issues of fairness and compliance, and ultimately leading to a loss of trust from users and decision-makers. Furthermore, these models make it difficult for engineers and data scientists to identify model deficiencies and edit decision-making or predictive behaviors.
  
  As a result, data analysts across various industries prefer interpretable and editable machine learning methods~\cite{vertsel2024hybrid,gerussi2022llm,Zhang2024LargeLM,truhn2023large}. The question then arises: how can PFMs aid in the development of interpretable and editable machine learning? The key lies in their potential to overcome current obstacles in explainable machine learning by leveraging their capacity to explore logical systems with greater expressiveness and human-understandable rules~\cite{reizingerposition}. By doing so, PFMs can facilitate the construction of models that are both interpretable and editable, aligning with the practical needs of data analysis. Despite the promising advantages, integrating PFMs to enhance interpretability and editability in machine learning introduces several challenges that must be addressed.
  
  \paragraph{Enhancing interpretability with high expressiveness.}
  
  There exists a fundamental trade-off between the expressiveness of a model and its interpretability. Highly expressive models, such as deep neural networks, can capture complex patterns within data but are often opaque, making them difficult for humans to understand. Conversely, interpretable models, like decision trees and rule-based systems, offer transparency but may lack the expressiveness required to model intricate data relationships.
  
  PFMs are adept at exploring logical systems with increased expressiveness while incorporating human-under\-standable rules~\cite{reizingerposition}. By leveraging PFMs, we can construct hypothesis spaces that balance expressiveness and interpretability, overcoming the traditional compromise between model complexity and understandability. One way to enhance interpretability without sacrificing expressiveness is to leverage PFMs to understand and interpret complex models for interpretable needs. \cite{nam2024using} propose an IDE plugin that can interpret complex code repositories whose code comments and documentation are scarce or hard to navigate. More pragmatically, analysts can distill PFM knowledge to simple and interpretable models~\cite{singh2023augmenting}. With constrained but enough expressiveness of simple models, analysts can make appropriate choices for consolidating augmented knowledge into simpler models.
  
  PFMs also help in interpreting models, and complex data structures, e.g., \cite{ko2024filling} designed an LLM-based method to convert scientific data into the desired form. Though a database with complete records is highly expressive in capturing reality with fixed, it is hard to interact with and explore. \cite{zheng2024revolutionizing} proposes a question-answering method for databases. This involves creating interfaces that transform data types challenging for humans to comprehend (e.g., continuous numerical values, extensive codebases) into more accessible forms (e.g., Boolean variables, symbolic expressions, natural language descriptions, visualizations). Utilizing PFMs to generate human-like explanations can achieve a level of diversity and clarity that approximates or even surpasses human capabilities.
  
  \paragraph{Enhancing editability with high stability.}
  
  Another challenge lies in balancing model editability with stability. Editable models allow engineers and data scientists to adjust and refine model behavior to meet new requirements, inject external knowledge, or eliminate undesired actions. However, frequent edits can lead to instability, potentially degrading the model's performance on previously generalized samples.
  
  Compared to uneditable neural network methods, PFMs, and editable machine learning models are poised to become new focal points~\cite{vojivr2020editable}. These models excel in adapting to new needs and eliminating unwanted behaviors. They naturally represent relational data, making them suitable for downstream tasks involving such data. For instance, using neural network models directly for classification tasks on tabular data often yields performance that is inferior to decision tree and rule-based methods~\cite{popov2019neural,grinsztajn2022tree}. Moreover, their interpretability and editability are limited, hindering the quick and accurate incorporation of new rules.
  
  By utilizing the rule extrapolation capabilities of PFMs~\cite{reizingerposition}, we can compensate for the shortcomings of these discriminative models. For example, \cite{nam2024optimized} employs pre-trained models to augment data for decision trees, enhancing their performance. From a practical standpoint, interpretable methods in tabular domains can be as accurate as black-box models. Researchers have leveraged interpretable machine learning to help practitioners identify defects in datasets, discover new scientific insights, and build fairer and more robust models~\cite{caruana2022data}.
  
  Data analysts prefer white-box models, therefore, researchers in business, medicine, energy, and other fields have begun to combine PFMs with interpretable models~\cite{vertsel2024hybrid,gerussi2022llm,Zhang2024LargeLM}. These methods excel in editability, carefully integrating PFMs' reasoning capabilities with editable machine learning algorithms to enhance model generalization without compromising performance on existing generalized samples.

  \subsection{PFMs Enhance Data Quality Optimization}\label{sec:data_quality}
  
  \begin{figure*}[h]
    \centering
    \includegraphics[width=0.7\textwidth]{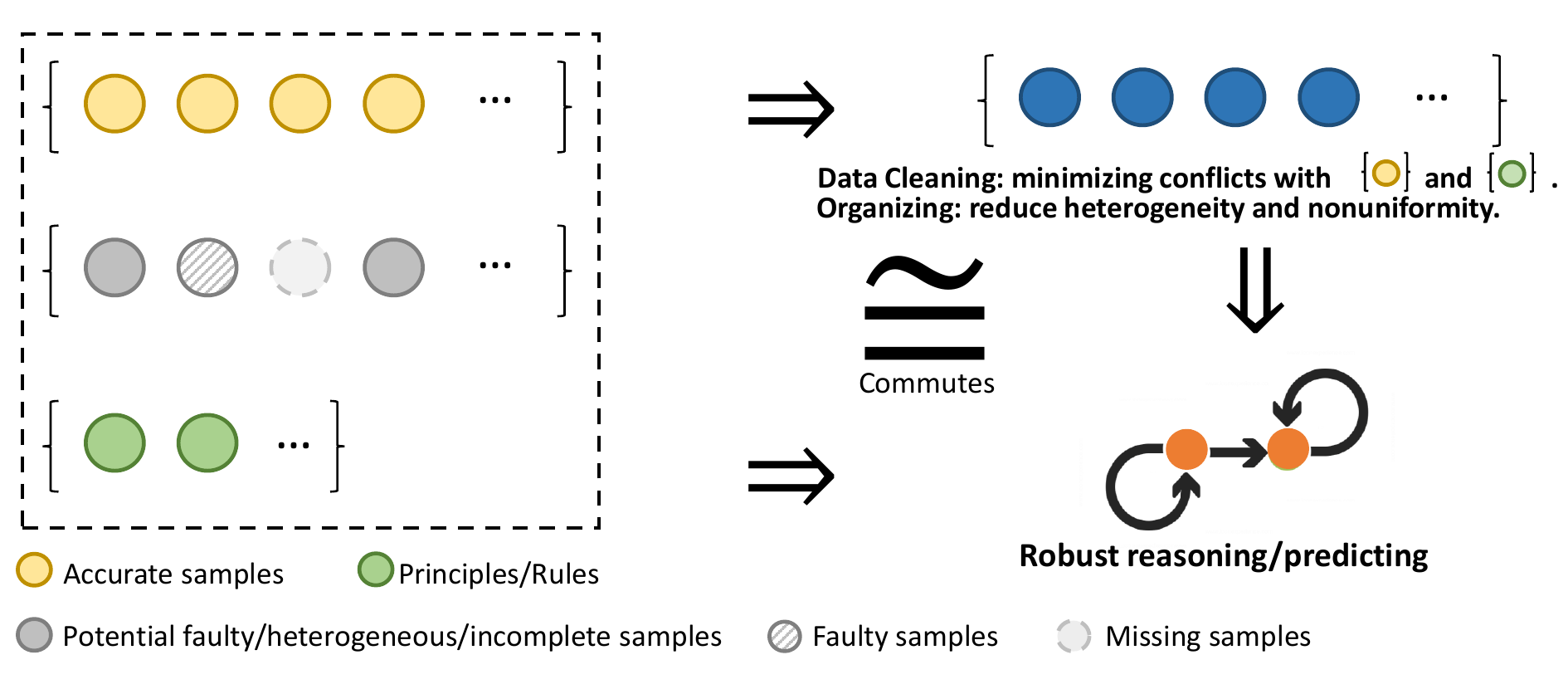} 
    \caption{\textbf{Overview of solving data quality issues.} This involves data preparation that can produce high-quality datasets or robust methods that directly produce models free from contradictions and incompleteness. High-quality data from data cleaning can be produced by coordinating with other datasets, external knowledge, and propositions but can introduce additional complexity for computation or management. Robust methods can be easy to apply but can introduce complexity to models and optimization methods. Idealistically these methods commute for the same data analysis task.}
    \label{fig:data_quality}
  \end{figure*}

  Data quality is a critical factor in the success of machine learning models and data analysis. Poor data quality can lead to incorrect conclusions, model overfitting, and reduced generalization ability. Pre-trained Foundation Models (PFMs) offer new avenues for enhancing data quality by addressing issues such as data incompleteness, uneven distribution, noise, and errors. By leveraging the strengths of PFMs in data generation, inference, and cleaning, we can improve the reliability and robustness of data analysis.
  
  \subsubsection{Representative Data Sampling/Generating}\label{sec:sample_generating}
  
  Representative sampling is essential to ensure that the data collected accurately reflects the underlying population or phenomenon. Beyond the concept of PAC learning, the distribution $\mathcal{D}$ from which the samples are drawn plays a crucial role in the learning process. If the sample does not represent $\mathcal{D}$ well, the learning algorithm $\mathcal{L}$ may not produce a hypothesis $h$ that approximates the true concept $c$ within the desired error bounds.
  
  \paragraph{Active Sampling.}
  
  PFMs can assist in active representative sampling by identifying regions in the data space where additional samples would most improve the model's performance. By doing so, they help mitigate issues related to insufficient or uneven data distributions, which are common challenges in real-world applications. For instance, when certain classes are underrepresented, PFMs can guide the collection of additional samples for those classes, enhancing the overall representativeness of the dataset. Solo~\cite{Wang2023SoloDD} and CHORUS~\cite{CHORUS} perform data discovery on large-scale databases to find helpful records from a large dataset, which shares the same goals of EDA with automation. 
  
  These techniques are always based on data mining methods such as schema matching~\cite{LLMschema} and entity resolution, to recognize stable patterns and linkages to compose complex semantics and search with. The identified patterns and reusable compositions construct the search space of sampling. ALT-GEN~\cite{tabulargeneration} benchmarks such ability of PFMs to search for possible unions among multiple tables. 
  
  \paragraph{Generating with Deductive Bias.}
  
  PFMs can generate synthetic data that adhere to known constraints and logical rules, a process we refer to as generating with deductive bias. Data generating involves generating representative entities and linkages according to these rules.

  For example, linkages would be missing in tabular data due to heterogeneity forms of the same entity. PFMs help to discover linkages to create more complete records to enhance semantic accuracy~\cite{Nobari2023DTTAE}. By leveraging prior knowledge and deductive reasoning, PFMs produce data that not only augment the existing dataset but also preserve essential properties and invariants of the data generation mechanism. This approach helps address data scarcity and imbalance by enriching the dataset with high-quality samples that reflect the underlying concepts more accurately.
  
  Generative models like generative adversarial methods~\cite{Loem2023SAIEFS}, and variational inference-based methods~\cite{huamortizing} enable PFMs to create samples similar to the training data with reorganized causes and effects inferred by these inference techniques implied in the data. By enhancing the model's understanding of data distributions and algebraic structures, these generative methods improve generalization ability and robustness. For example, PFMs can generate data beneficial to analysis through adversarial generation~\cite{du2024enhancing,weng2023g}, effectively capturing global characteristics of the data.
  
  Researchers have found that PFMs can enhance Bay\-esian inference and variational inference methods~\cite{huamortizing}. These methods infer important structures within the data and generate new data based on these structures, significantly improving the accuracy and reliability of data analysis in scientific research and industrial applications.
  
  \subsubsection{Robustness Against Errors}\label{sec:robustness}
  
  Robustness against errors is vital for the reliability of machine learning models. Errors in data, such as noise, missing values, and inconsistencies, can lead to overfitting and reduced model performance. In the PAC learning framework, robustness can be viewed as the ability of the learning algorithm $\mathcal{L}$ to produce a hypothesis $h$ that approximates the true concept $c$ even when the data distribution $\mathcal{D}$ contains errors.
  
  \paragraph{Data Cleaning.}
  
  PFMs play a significant role in data cleaning by detecting and correcting errors in datasets. They can identify data points that violate known data quality rules and constraints~\cite{li2024towards}, assisting analysts in cleaning the data effectively. By automating error detection and correction, PFMs improve the overall data quality, which is essential for reliable inductive reasoning as discussed earlier.
  
  Data analysts face challenges in accurately and efficiently representing domain or expert knowledge using predictive models and domain-specific languages. The optimization of data quality is essentially about using limited data, combined with expert knowledge, to perform induction and deduction in the hypothesis space $\mathcal{H}$, continuously providing conservative rules closer to the data generation mechanism~\cite{peng2022self}. PFMs aid in this process by offering implicit representations of potentially unknown information, serving as a valuable starting point for data quality optimization.
  
  \paragraph{Robust Methods.}
  
  Developing robust hypothesis classes and learning algorithms is crucial for handling errors and noise in data. PFMs contribute to the creation of robust methods by enhancing the model's capacity to generalize from imperfect data. For example, PFMs can help in constructing hypothesis spaces that include invariants and constraints derived from algebraic~\cite{Miao2022LearningIA}, analytic~\cite{Neu2022GeneralizationBV}, and geometric methods~\cite{Atzeni2023InfusingLS}, which improve the model's generalization ability even when data quality is compromised.
  
  Moreover, generative models have significant advantages in addressing data missing and uneven distribution issues. By learning the potential distribution of the data, generative models can generate new samples that better capture the global characteristics of the data~\cite{094dai2017good}. Studies have shown that in low-resource environments, the performance of generative models surpasses that of discriminative models~\cite{097yoon2017semi}. This is primarily because generative models can generate more high-quality training data to compensate for the lack of labeled data, thereby improving the model's generalization ability. Discriminative models, which rely directly on existing labeled data, cannot fully utilize the potential information in unlabeled data~\cite{102yang2018deep}.
  
  By integrating PFMs into the data quality optimization process, we can enhance the robustness and reliability of machine learning models, ensuring that they perform effectively even in the presence of data imperfections. This aligns with the principles of PAC learning, where the goal is to learn approximately correct concepts from data distributions, and supports both inductive and deductive reasoning in the analysis.
  
  \subsection{Automated Machine Learning}\label{sec:auto_ml}
  
  Automated machine learning (AutoML) aims to automate the design, deployment, and optimization processes of machine learning models, enabling non-domain experts to effectively utilize data and reducing the need for manual intervention by professional data scientists. It includes model selection and optimization, feature engineering~\cite{Hollmann2023LargeLM}, hyperparameter tuning, model evaluation, and end-to-end process automation~\cite{salehin2024automl}. The following definition and insight highlight the importance of choosing appropriately represented sample distributions, learning algorithms, and hypothesis spaces.
  
  \paragraph{Definition.} Given a sample size \\ $m \geq \text{poly}(1/\epsilon, 1/ \delta, \text{size}(x), \text{size}(c))$, if a learning algorithm $\mathcal{L}$ makes the concept class $\mathcal{C}$ PAC identifiable, and the running time of $\mathcal{L}$ is a polynomial function $\text{poly}(1/\epsilon, 1/ \delta, \text{size}(x), \text{size}(c))$, then the concept class $\mathcal{C}$ is efficiently PAC learnable. Here, $\text{size}(x)$ and $\text{size}(c)$ represent the complexity or dimensionality of a single sample and a single concept, respectively.
  
  \paragraph{Insight.} In machine learning, for the same concept class, choosing different representations, learning algorithms, and hypothesis spaces can lead to different learnability outcomes.

  \subsubsection{Consolidating AutoML}\label{sec:consolidate_automl}
  
  Automated Machine Learning (AutoML) aims to automate the design, deployment, and optimization processes of machine learning models, enabling non-domain experts to effectively utilize data and reducing the need for manual intervention by professional data scientists. It's an inquiry for automation of self-disciplined reasoning and decision plans with the least human interference.
  
  However, constructing AutoML projects often requires significant manual effort in configuring pipelines, selecting appropriate algorithms, and tuning parameters. These are efforts to consolidate machine learning pipelines to where more effective hypotheses lie. Furthermore, researchers are exploring more effective learning methods compatible with increasingly complex search spaces. The choice of representations (controlled by the form of selected feature), learning algorithms (partially tuned by hyperparameter), and hypothesis spaces (controlled by model architecture) significantly affect the identifiability of a concept class.
  
  Recent works have utilized PFMs to enhance these various aspects of AutoML. For instance,~\cite{Hollmann2023LargeLM} achieved automatic feature engineering based on large language models (LLMs), streamlining the process of extracting relevant features from data. Additionally,\cite{sayed2024gizaml} employed PFMs to explore neural architecture search and hyperparameter spaces for foundational learning and inference methods. These approaches leverage the capabilities of PFMs to navigate complex search spaces efficiently.
  
  Pre-trained Foundation Models (PFMs) have demonstrated great potential in automating model selection and optimization within the data analysis process~\cite{liu2023jarvix}. Recent studies have shown that PFMs can implement standard machine learning algorithms in different contexts, such as least squares regression and gradient descent for two-layer neural networks~\cite{bai2024transformers}. This suggests that PFMs can implicitly perform algorithm selection and optimization, reducing the manual effort required in constructing AutoML projects.

  \subsubsection{Scaling AutoML}\label{sec:scaling_automl}
  
  To move beyond FE, NAS, and HPO, recent research focuses on incorporating PFMs into other stages of the machine learning pipeline, such as data augmentation, model ensembling, and transfer learning. This involves more explicit management of formal knowledge which encodes consistent structural methods~\cite{SongY00024, HsuMTW23}, and is discussed more elaborately in \S~\ref{sec:dsl}. This holistic integration can enhance the overall performance and scalability of AutoML systems, enabling them to handle a wider range of tasks and datasets with minimal human intervention.
  
  The application of PFMs in context learning provides a new perspective for algorithm selection. Models like GPT and BERT are capable of understanding and processing complex contextual information, and based on this, selecting appropriate algorithms~\cite{009brown2020language, EoTGE}. These models leverage simple underlying rules, and as pointed out by~\cite{reizingerposition}, PFMs exhibit near-saturated statistical generalization properties and possess certain rule extrapolation capabilities. This endows them with strong context learning abilities, enabling them to adaptively select and optimize algorithms. Moreover, these extrapolation rules can be explicitly expressed as formal languages by PFMs~\cite{cheng2022binding}, enhancing the interpretability of AutoML results.
  
  Furthermore, transducing the knowledge searched during the AutoML process into understandable formats is facilitated by PFMs' ability to generate explanations and formal representations. This enhances the interpretability of the models and allows analysts to gain insights into the reasoning behind model selections and predictions.
  
  \section{Challenges}
  
  The rise of pre-trained foundation models (PFMs), such as GPT-3 and DALL·E, has brought significant transformations to the field of data analysis. By integrating various structured and semi-structured data (relational data, time-series data, text, images, audio, etc.) and employing methods that allow for formal verification to mine invariants and relationships, they provide more comprehensive and complex data analysis capabilities. However, despite their immense potential in certain applications, there are still some shortcomings in their application to data analysis.
  
  \paragraph{Optimization of Inference Costs.} The efficiency issue of large models is a significant challenge. These models typically require substantial computational resources for inference and training, leading to high costs and scalability difficulties in practical applications~\cite{041zhai2022scaling}. \cite{chow2024performance} studied the costs of large models, estimating that operating ChatGPT costs over \$700,000 per day, and small businesses using GPT-4 to support customer service may incur monthly costs exceeding \$21,000. The high infrastructure and financial costs, coupled with the need for specialized talent, make LLM technology unattainable for most organizations. Additionally, the upfront costs of using this technology include emissions from manufacturing the related hardware and the costs of running that hardware during training.
  
  \paragraph{Domain Generalization Ability.} Therefore, rapidly adapting large models and consolidating their reasoning and learning capabilities have become an urgent problem to solve. Many pre-trained models exhibit poor transferability between different tasks, requiring frequent adjustments and fine-tuning. Applying pre-trained models to different datasets and tasks usually necessitates substantial reconfiguration, with a variety of methods including prompt engineering~\cite{sordoni2024joint}, retrieval-augmented generation (RAG)~\cite{gao2023retrieval}, fine-tuning, knowledge editing~\cite{Wang2023KnowledgeEF}, reinforcement learning~\cite{Kaufmann2023ASO}, and incremental learning~\cite{wu2024continual}. The rapid iteration of technology increases development and maintenance costs and affects the speed of initiating data analysis tasks. Although these technologies claim to achieve domain transfer of large models at lower costs, meticulous and tedious adjustments are still required for each new task. From a data analysis perspective, it is essential to discuss the principles for selecting domain transfer technologies, helping researchers clarify the use of these techniques to approach best practices.
  
  \paragraph{Limitations of Consistency.} Moreover, these technologies have not currently solved the fundamental limitations of large models~\cite{bender2021dangers}. For example, issues of bias in large models, hallucinations~\cite{Ji2022SurveyOH}, lack of robustness to different forms of prompts~\cite{Lu2021FantasticallyOP}, and susceptibility to interference from irrelevant content in prompts and training data~\cite{Shi2023LargeLM}. Researchers should learn to coexist with these inherent limitations of large models when handling data analysis tasks. Similarly, from a data analysis perspective, providing a roadmap to understand and overcome these limitations is crucial.
  
  \paragraph{Application of Fundamental Theories and Tools.} Knowledge being treated with disdain may be due to the maturity of the field; furthermore, pre-trained models in data analysis often rely on new tools and techniques, which may lead to the neglect of the importance of traditional theories, constituting a risk of knowledge regression. For example, optimization theory and relational data models~\cite{codd2007relational} are core theories in data analysis. Therefore, it is necessary to use these foundational methods to mutually promote applications of large models~\cite{li2024can}. The relational data model has irreplaceable advantages in handling structured data, but in the application of multimodal models, it is often supplanted by more novel techniques, leading to a disconnect between theory and practical applications~\cite{dinh2024large}. To reduce data redundancy and anomalies, the relational dependency paradigm remains crucial, while large models often lack effective methods when dealing with these issues. Additionally, directly applying large models to a specific type of data and task is often undesirable~\cite{tan2024language}, indicating that large models are not a panacea for all problems.
  
  \paragraph{Trust Issues.} Explainability aims to use large models to obtain trustworthy analytical conclusions and conservatively seek more optimized analytical methods. However, the reasoning of large models is accompanied by uncertainty and the inherent ambiguity of natural language. Therefore, seeking stronger functional dependencies, such as provable theorems or other logical forms, is essential~\cite{morishita2023learning,pei2023can,abbe2023generalization,yang2024leandojo}. Combining large models with traditional data analysis theories can enhance the reliability and interpretability of models~\cite{khakhar2023pac}. The integration of foundational data analysis methods with pre-trained models can better solve complex data analysis problems.
  
  Although large multimodal pre-trained models have shown great potential in data analysis, issues such as their cost, maintainability, interpretability, robustness and reliability, and compatibility with foundational theories still need further resolution. These constitute significant challenges in further applying PFMs to data analysis.

   \section{Future Research Outlook}
  
  After a comprehensive examination of the current state of data analysis, future research directions can be envisioned from the following three dimensions.
  
  \subsection{Further Deepening and Developing Existing Data Analysis Methods}
  
  Future research will continue to deepen existing methods, including the further refinement of relational data models, adaptive improvements of decision tree algorithms for different data distributions, performance optimization of clustering analysis when handling large-scale datasets, innovation in artificial neural network architectures, and intelligent upgrades of automated machine learning technologies. Additionally, data augmentation methods based on rules or closed-domain data will be a research focus, aiming to enhance model generalization and adaptability to specific domain data.
  
  \subsection{Data Analysis Methods Enhanced by Pre-trained Foundation Models}
  
  The introduction of pre-trained foundation models (LMs) has brought new perspectives to data analysis. Future research will explore how to enhance models' understanding of complex data through knowledge-augmented neural networks and how to utilize generative open-domain data augmentation techniques to expand datasets and improve model generalization. Meanwhile, research on complex prediction and inference methods will assist in solving more intricate data analysis problems, such as time-series forecasting and causal inference.
  
  \subsection{New Possibilities Brought by Pre-trained Foundation Models}
  
  LM technology will usher in a new chapter in data analysis. Researchers will explore automatic code generation and optimization techniques to reduce development time and improve code quality. Automatic goal planning, as well as automated mathematical discovery and theorem proving, may change how we solve complex problems. Multilevel modeling of complex systems will help us understand and predict system behaviors more comprehensively. Rapid domain adaptation and automatic intervention with counterfactual reasoning technologies will provide more precise support for decision-making.
  
  With continuous technological advancements, future data analysis will become more intelligent and automated, capable of handling more complex data and problems. Researchers need to continuously explore new algorithms, models, and application scenarios to promote sustained development and innovation in the field of data analysis.
  
  \section{Conclusion}
  
  Through an in-depth revisitation of data analysis, we recognize that PFMs can systematically optimize the data analysis process. These models, utilizing deep learning and neural network-based methods, not only improve the efficiency of data processing but also enhance the ability to extract valuable information from data. It has demonstrated significant advantages in handling large-scale and complex datasets, inductive, and deductive reasoning, and identification of new concepts.
  
  We have also revealed universal challenges faced in the application prospects of large model technology in various fields, especially in business, finance, and healthcare. We demonstrate that large model technology has the potential to greatly improve the quality and efficiency of decision-making. Future research will further explore the application effects of these models in specific domains and how to enhance their practicality and accuracy through continuous model optimization and iteration.

  However, we can also observe the challenges that large model technology faces in practical applications, including high computational resource consumption. These issues suggest that future research needs to find a balance between technological innovation and existing challenges to promote the sustainable development of data analysis technology.

  In summary, the future development of data analysis will be multifaceted, requiring comprehensive consideration of technological innovation, practical applications, ethical regulations, and social impact. Through a comprehensive analysis of existing literature, we have reason to believe that large model technology will continue to serve as an important driving force in the field of data analysis, promoting continuous progress and innovation in related technologies.

  
  \printbibliography

  \end{document}